\documentclass[aps,twocolumn,pre,floatfix]{revtex4-2}

\usepackage{xcolor,hyperref}
\hypersetup{
   colorlinks,
   linkcolor={blue!50!black},
   citecolor={blue!50!black},
   urlcolor={blue!80!black}
}

\usepackage{epsfig, amsmath}
\usepackage{bm}
\usepackage{soul,ulem}
\usepackage{hyperref}
\usepackage{footmisc}
\usepackage{wrapfig}
\DeclareMathAlphabet{\mathitbf}{OML}{cmm}{b}{it}

\renewcommand{\=}{\!=\!}
\DeclareMathAlphabet\mathbfcal{OMS}{cmsy}{b}{n}

\begin{document}

\title{An equation of motion for unsteady frictional slip pulses}
\author{Eran Bouchbinder}
\affiliation{Chemical and Biological Physics Department, Weizmann Institute of Science, Rehovot 7610001, Israel}
\begin{abstract}
Frictional sliding, e.g., earthquakes along geological faults, are mediated either by frictional crack-like ruptures, where interfacial (fault) slip is accumulated during the entire sliding event, or by frictional pulse-like ruptures, featuring a finite length over which slip is accumulated. Our basic understanding of slip pulses, which are believed to dominate most crustal earthquakes, is still incomplete. Here, building on recent progress, we present an analytic equation of motion for rate-and-state frictional slip pulses, which are intrinsically unstable spatiotemporal objects, in terms of a single degree of freedom. The predictions of the equation are supported by large-scale simulations of growing pulses and reveal the origin of the slow development of their instability, which explains the dynamic relevance of pulses in a broad range of natural and manmade frictional systems.
\end{abstract}

\maketitle

{\em Introduction}.---Sliding along interfaces between materials in frictional contact, e.g., earthquakes along geological faults, is mediated by the propagation of inhomogeneous spatiotemporal modes. These are generally classified into expanding crack-like modes, where interfacial slip is accumulated during the entire sliding event, and pulse-like modes, which are compact objects featuring a finite length~\cite{freund1979mechanics,Heaton1990,Perrin1995,Beroza_Mikumo_1996,Beeler1996,Cochard1996,Andrews1997,Zheng1998,Nielsen2000,Nielsen2003,Brener2005,Shi2008,Rubin2009,Garagash2012,platt2015steadily,Gabriel2012,Putelat2017,Michel2017,lambert2021propagation,ROCH2022104607,galetzka2015slip,Dunham2011a,Brener2018,pomyalov2023self,brantut2019stability,pomyalov2023dynamics,pomyalov2024,gabrieli2025lab}. The latter implies that slip at a fixed location on the interface/fault is accumulated over a finite time, which is typically much shorter than the event's duration. It is believed that slip pulses are the dominant rupture modes in most crustal earthquakes, and likely also in many tribological systems.

Our current understanding of slip pulses lags behind that of crack-like modes. The origin of the difficulty is two-fold. First, crack-like frictional rupture bears some analogies to ordinary cracks in bulk materials~\cite{Svetlizky2019,Barras2019,Barras2020}, which are better understood, while no analogy to slip pulses in bulk materials exists. The deep reason for the latter is the crux of the second origin of difficulty; slip pulses involve the spatiotemporal interaction between two propagating fronts --- a rupture (leading) front and a healing (trailing) front --- which self-consistently selects the pulse length, propagation velocity and total accumulated slip. Healing processes, which lead to interfacial strengthening and re-locking, are absent in bulk materials and are inherently related to the dynamics of the contact interface~\cite{Dieterich1978,Dieterich1994a,Beeler1994,Marone1998a,Berthoud1999,Baumberger2006Solid,Ben-David2010}.

At present, there exists no well-established and extensively validated equation of motion for frictional slip pulses. Yet, recent progress provided deep insights into the properties and structure of such an equation~\cite{Brener2018,brantut2019stability,pomyalov2023self,pomyalov2023dynamics}. First, it has been shown that slip pulses are intrinsically unstable objects~\cite{Perrin1995,Brener2018,brantut2019stability,pomyalov2023dynamics}, i.e., they either decay or grow, corresponding to a ``saddle configuration'' in a non-equilibrium analogy to critical nuclei in equilibrium first-order phase transitions~\cite{Brener2018,pomyalov2023dynamics}. Second, it has been shown that unsteady slip pulses are intrinsically related a family of steady-state pulse solutions~\cite{brantut2019stability,pomyalov2023dynamics}; specifically, an unsteady slip pulse at a given driving stress --- whether decaying or growing --- evolves as a continuous sequence of steady-state pulses corresponding to different driving stresses. Finally, it has been shown that the instability evolution of growing pulses is slow such that they propagate many times their characteristic size without appreciably changing their properties~\cite{brantut2019stability,pomyalov2023dynamics}.

Based on these new insights and properties, an approximate equation of motion has been recently proposed~\cite{brantut2019stability}. The equation has been shown to agree with the early-time evolution of the instability of steady-state slip pulses driven by thermal pressurization of pore fluids within the fault~\cite{Rice2006,Viesca2015,brantut2019stability}. Here, building on these recent developments, we present an analytic equation of motion for slip pulses obtained for a generic rate-and-state friction constitutive law~\cite{BarSinai2012,Bar-Sinai2014,Brener2018,Bar-Sinai2019,Barras2019,Barras2020,pomyalov2023self,pomyalov2023dynamics,pomyalov2024}, which is very different from the thermal pressurization one. The predictions are supported by large-scale simulations of growing pulses, both near and far from the steady-state point, and reveal the origin of the slow development of their instability. Overall, our results support and substantiate the validity of the emerging equation of motion for frictional slip pulses.
\begin{figure}[htbp!]
    \centering
    \includegraphics[width=1\linewidth]{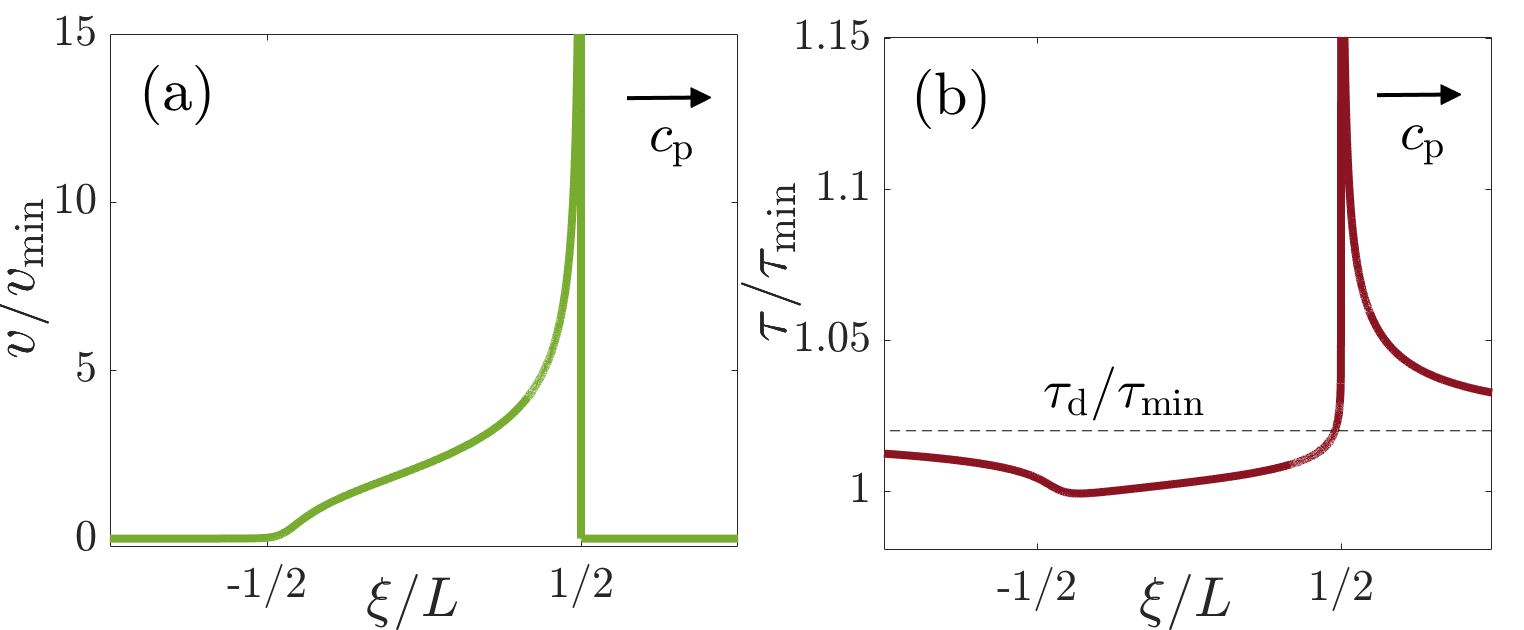}
	\caption{An example of a steady-state rate-and-state friction slip pulse, obtained in~\cite{pomyalov2023self} with $\tau_{\rm d}\!=\!1.02\tau_{\rm min}$. The pulse, of size $L$, propagates from left to right at a velocity $c_{\rm p}$. Shown are the slip velocity field $v(\xi)$ (panel a) and the shear stress $\tau(\xi)$ (panel b, the driving stress $\tau_{\rm d}$ is marked by the horizontal dashed line), where $\xi\!=\!x\!-\!c_{\rm p}t$ is a co-moving coordinate. The steady-state friction curve $\tau_{\rm ss}(v)$ in~\cite{pomyalov2023self} attains a local minimum at $\tau_{\rm ss}(v_{\rm min})\!=\!\tau_{\rm min}$, see text for discussion.}
    \label{fig:pulse_illustration}
\end{figure}

{\em Effective mapping to ideal pulses}.---As mentioned above, and as will be further discussed below, there exist fundamental relations between the unsteady dynamics of slip pulses under a fixed background/driving stress $\tau_{\rm d}$ and a family of steady-state solutions parameterized by varying $\tau_{\rm d}$. Consequently, we first develop an analytic description of steady-state slip pulse solutions obtained under rate-and-state friction~\cite{pomyalov2023self}. The latter constitutive framework expresses the frictional strength $\tau(x,t)$ as a functional of the slip velocity $v(x,t)$ and a structural (``internal'') state field $\phi(x,t)$, which is related to the real contact area along the frictional interface~\cite{Marone1998a,Berthoud1999,Baumberger2006Solid}. Here, $x$ is the coordinate along the contact interface residing at $y\=0$ between two semi-infinite linear elastic bodies, and $t$ is time.

The displacement field $u_z(x,y,t)$, where $z$ is perpendicular to both $x$ and $y$ (the so-called mode-III symmetry), satisfies a scalar wave equation $\mu {\bm \nabla}^2 u_z\=\rho\partial_{tt}u_z$, where $\mu$ is the shear modulus and $\rho$ is the mass density. The driving stress $\tau_{\rm d}$ is applied anti-symmetrically at $y\!\to\!\pm\infty$ along with a compressive (normal) stress that keeps the bodies in contact. The slip displacement is $\delta(x,t)\!\equiv\!2u_z(x,y\=0,t)$, the slip velocity is $v(x,t)\=\partial_t\delta(x,t)$ and the interfacial boundary condition reads $\tau[v(x,t),\phi(x,t)]\=\sigma_{yz}(x,y\=0,t)\=\mu\partial_y u_z(x,y\=0,t)$. $\phi(x,t)$ satisfies its own time evolution equation that accounts for contact rejuvenation under sliding and contact aging/healing in the absence of sliding. Under steady sliding at a slip velocity $v$, the latter two physical processes balance each other and $\phi$ is expressed in terms of $v$~\cite{Marone1998a,Berthoud1999,Baumberger2006Solid}, resulting is a steady-state friction curve $\tau_{\rm ss}(v)$. The latter has been shown by numerous experiments to be N-shaped~\cite{Bar-Sinai2014}, attaining a local minimum at $\tau_{\rm ss}(v_{\rm min})\=\tau_{\rm min}$.

Steady-state pulse solutions are expressed in terms of the co-moving coordinate $\xi\=x\!-\!c_{\rm p}t$, where $c_{\rm p}\=\beta c_s$ is an unknown pulse propagation velocity ($c_s\=\sqrt{\mu/\rho}$ is the shear wave-speed), with the additional boundary condition $v(\xi\!\to\!\pm\infty)\=0$. The latter distinguishes crack-like from pulse-like solutions as it demands that $v$ vanishes both ahead of the leading rupture edge and behind its trailing (healing) edge, thus selecting both the unknown pulse size $L$ and the dimensionless propagation velocity $\beta$. An example with $\tau_{\rm d}\=1.02\tau_{\rm min}$, obtained numerically in~\cite{pomyalov2023self}, is presented in Fig.~\ref{fig:pulse_illustration}. A continuous family of such steady-state rate-and-state friction slip pulses as a function of $\tau_{\rm d}$ has been obtained numerically in~\cite{pomyalov2023self}, but we lack an analytic description of such solutions, a problem addressed next.

In the pulse solutions described above and demonstrated in Fig.~\ref{fig:pulse_illustration}, $\beta$, $L$ and the stress distribution inside $\tau(\xi)$ inside the pulse $|\xi|\!\le\!L/2$ are self-consistently selected. If, on the other, a constant residual stress $\tau_{\rm res}\!<\!\tau_{\rm d}$ is assumed for $|\xi|\!\le\!L/2$ as an input, along with $v(|\xi|\!>\!L/2)\=0$ (no sliding out of the pulse) and additional constraints on the behavior at the trailing/healing ($\xi\=-L/2$) and leading ($\xi\=L/2$) edges are posed, an analytic solution can be obtained. Specifically, demanding continuous fields at $\xi\=-L/2$ and allowing singular/divergent fields at $\xi\=L/2$, yet accompanied by a finite energy flux that is dissipated per unit propagation --- by amount $G_{\rm c}$ --- at the singularity, the solution $v(\xi)\= v_0 \sqrt{(\tfrac{1}{2}L+\xi)/(\tfrac{1}{2}L-\xi)}$ for $|\xi|\!\le\!L/2$ is obtained~\cite{freund1979mechanics}. We term it hereafter the ideal steady-state pulse solution.

The ideal pulse solution contains two input parameters, $\tau_{\rm res}$ and $G_{\rm c}$ (both assumed to be known), $\tau_{\rm d}$ as a control parameter, and three unknowns, $\beta$, $L$ and $v_0$, which should be computed. As will become clear below, it would be useful to relate $v_0$ to the total accumulated slip $b$, given as $b\=(\beta c_s)^{-1}\!\!\int_{-L/2}^{L/2} v(\xi)d\xi$. The latter implies $v_0\=2\beta c_s b/(\pi L)$. To determine the 3 unknowns, we need 3 relations. One corresponds to the condition of continuous fields at $\xi\=-L/2$, taking the form $\pi L(\tau_{\rm d}-\tau_{\rm res})\=\mu\,b \sqrt{1-\beta^2}$, where $\tau_{\rm d}\!-\!\tau_{\rm res}$ is the stress drop accompanying pulse propagation. Another relation corresponds to energy balance at the leading edge, taking the form $G_{\rm c}\=(\tau_{\rm d}-\tau_{\rm res})b$. However, there exist no other relations, implying that the ideal pulse solution of~\cite{freund1979mechanics} is incomplete, i.e., some extra physics is missing.
\begin{figure}[htbp!]
    \centering
    \includegraphics[width=0.85\linewidth]{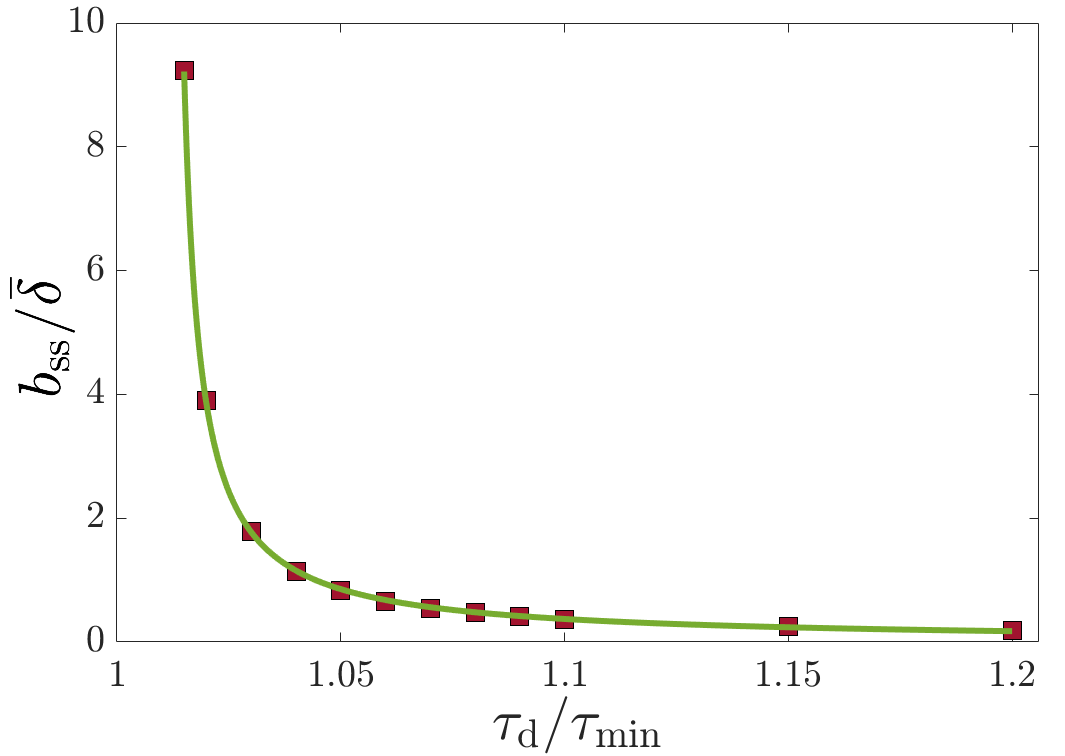}
	\caption{Testing the prediction for $b_{\rm ss}(\tau_{\rm d})$ in Eq.~\eqref{eq:b_ss_vs_tau_d} (solid line) against the numerical data of~\cite{pomyalov2023self} (squares). Excellent agreement with $G_{\rm c}\!=\!0.51$ J/m$^2$ and $\tau_{\rm res}\!=\!1.0114\tau_{\rm min}$ is demonstrated (the value of $\tau_{\rm min}$ is given in~\cite{SM} and the characteristic accumulated slip $\bar{\delta}$ is defined in the text).}
    \label{fig:b_ss_vs_tau_d}
\end{figure}

The missing physics emerges from the interfacial constitutive law, which both provides the extra relation and at the same time allows to effectively map the steady-state rate-and-state friction solutions illustrated in Fig.~\ref{fig:pulse_illustration} to the analytic ideal pulse solutions parameterized by $\tau_{\rm d}$. The important point is that the local minimum of the steady-state rate-and-state friction curve discussed above, obtained at $\tau_{\rm ss}(v_{\rm min})\=\tau_{\rm min}$, provides characteristic slip velocity and stress levels inside the pulse, as seen in Fig.~\ref{fig:pulse_illustration}, for driving stresses $\tau_{\rm d}$ not much larger than $\tau_{\rm min}$. Specifically, we expect $v_{\rm min}$ to be approximately determined by $v_0$ and $\tau_{\rm min}$ by $\tau_{\rm res}$ in the ideal pulse solutions, approximately independently of $\tau_{\rm d}$ not much larger than $\tau_{\rm min}$.

To test this expectation, we use the above-discussed ideal pulse solution to obtain the following relations
\begin{eqnarray}
    \label{eq:b_ss_vs_tau_d}
    b_{\rm ss}(\tau_{\rm d}) &=& \frac{G_{\rm c}}{\tau_{\rm d}-\tau_{\rm res}} \ ,\\
    \label{eq:L_ss_vs_beta_ss}
    L_{\rm ss}(\beta_{\rm ss}) &=& \frac{4}{\pi} \left(\frac{c_s}{v_0}\right)^{\!2}\,\frac{G_{\rm c}}{\mu}\frac{\beta^2_{\rm ss}}{\sqrt{1-\beta^2_{\rm ss}}} \ ,
\end{eqnarray}
where the subscript `ss' highlights the steady-state nature of these relations. We test the validity of Eq.~\eqref{eq:b_ss_vs_tau_d}, which does not involve $v_0$, against the pulse solutions of~\cite{pomyalov2023self}, parameterized by $\tau_{\rm d}$, in Fig.~\ref{fig:b_ss_vs_tau_d}. An excellent agreement is obtained, with $G_{\rm c}\=0.51$  J/m$^2$ (consistently with independent earlier estimates~\cite{Barras2020,pomyalov2023self}) and $\tau_{\rm res}\!=\!1.0114\tau_{\rm min}$, in line with the above expectation. Quite remarkably, the relation holds for all values of $\tau_{\rm d}$, also away from $\tau_{\rm min}$. We hereafter use the determined values of $G_{\rm c}$ and $\tau_{\rm res}$.

\begin{figure}[htbp!]
    \centering
    \includegraphics[width=1\linewidth]{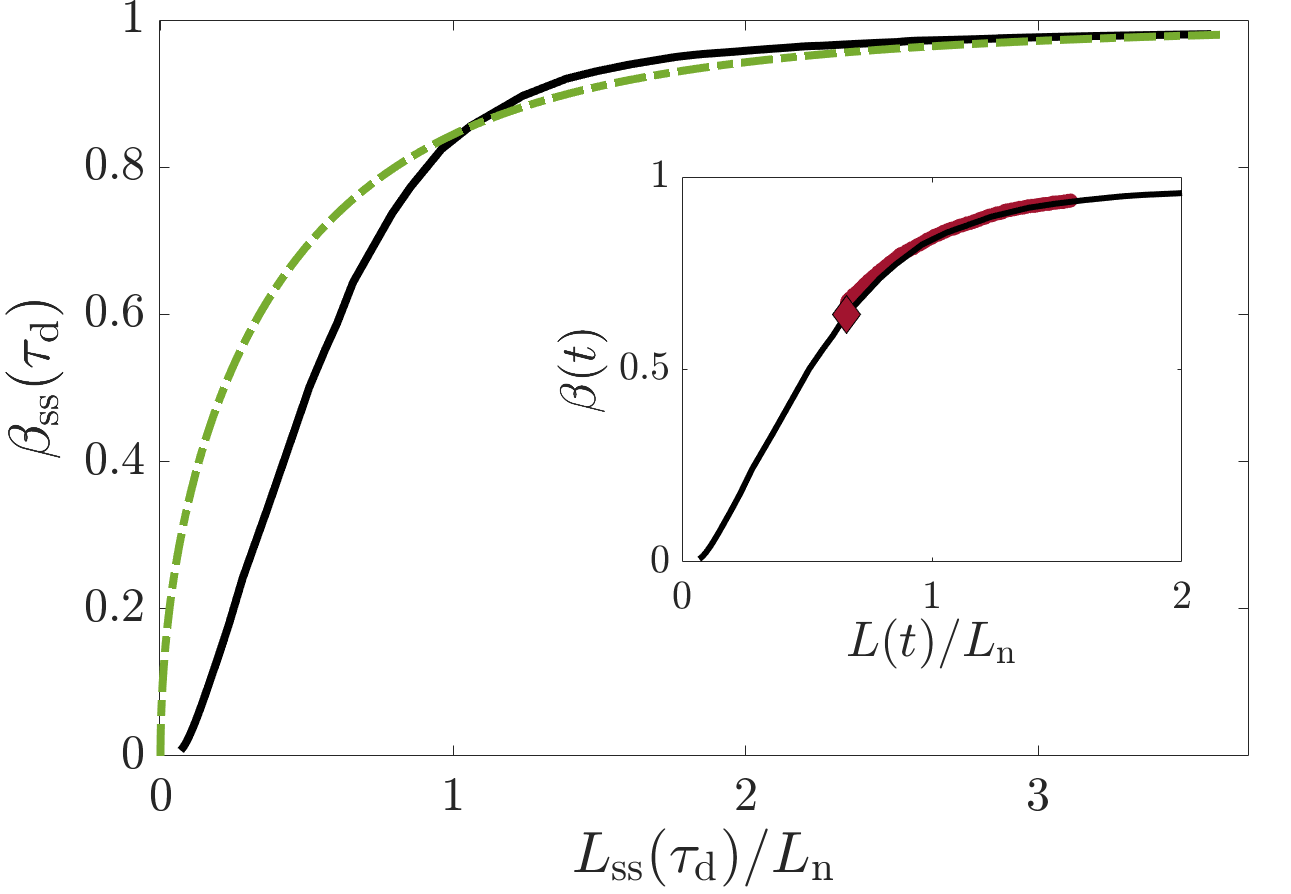}
	\caption{(main) Testing the prediction for $\beta_{\rm ss}(L_{\rm ss})$ (dashed line) in Eq.~\eqref{eq:L_ss_vs_beta_ss} against the numerical data of~\cite{pomyalov2023self} (solid line, parameterized by $\tau_{\rm d}$) using $v_0\!=\!1.15v_{\rm min}$. The known parameters $G_{\rm c}\!=\!0.51$ J/m$^2$, $c_s\!=\!2739$ m/s, $\mu\!=\!9$ GPa and $L_{\rm n}\!=\!15.2$ m were used~\cite{SM}. (inset) The time-dependent dynamics of a growing pulse (thick brown line) in the $\beta(t)\!-\!L(t)$ plane for a fixed $\tau_{\rm d}\!=\!1.04\tau_{\rm min}$, initiated near the steady-state point (large diamond, see also Fig.~3 of~\cite{pomyalov2023dynamics}). The steady-state (solid black) line is the same as in the main panel.}
    \label{fig:L_ss_vs_beta_ss}
\end{figure}
Next, we test the validity of Eq.~\eqref{eq:L_ss_vs_beta_ss} against the pulse solutions of~\cite{pomyalov2023self}. That is, we ask whether the rate-and-state slip pulse length-velocity $L_{\rm ss}(\beta_{\rm ss})$ relation is approximately described by Eq.~\eqref{eq:L_ss_vs_beta_ss} with $v_0\!\simeq\!v_{\rm min}$, at least for $\tau_{\rm d}$ values not far away from $\tau_{\rm min}$. The results are presented in Fig.~\ref{fig:L_ss_vs_beta_ss} (main panel), where the numerical rate-and-state $\beta_{\rm ss}(L_{\rm ss})$ data appear in the solid line and Eq.~\eqref{eq:L_ss_vs_beta_ss} with $v_0\=1.15v_{\rm min}$ corresponds to the dashed-dotted line. The latter value, to be used hereafter, is selected to match the data in the large $L_{\rm ss}$ and $\beta_{\rm ss}$ regime, corresponding the small $\tau_{\rm d}$~\cite{pomyalov2023self}. The agreement is reasonable and the results to follow do not depend on the exact value of $v_0\!\simeq\!v_{\rm min}$. Consequently, Eqs.~\eqref{eq:b_ss_vs_tau_d}-\eqref{eq:L_ss_vs_beta_ss} offer an analytic description of the steady-state rate-and-state friction pulses of~\cite{pomyalov2023self}. Additional steady-state relations, e.g., $\beta_{\rm ss}(b_{\rm ss})$ to be used below, can be readily derived~\cite{SM}.

{\em An analytic equation of motion for unsteady rate-and-state slip pulses}.---As mentioned above, it has been recently shown that steady-state pulse solutions, for any $\tau_{\rm d}$, are intrinsically unstable~\cite{Brener2018,pomyalov2023dynamics}. That is, pulses slightly larger than $L_{\rm ss}(\beta_{\rm ss})$ grow and those slightly smaller decay. Nevertheless, it has been shown that the family of steady-state pulse solutions is fundamentally related to the unsteady spatiotemporal dynamics of slip pulses at a fixed $\tau_{\rm d}$. Specifically, both growing and decaying pulses follow the steady-state curve formed by any pair of pulse observables, parameterized by $\tau_{\rm d}$. An example in the $L_{\rm ss}(\tau_{\rm d})\!-\!\beta_{\rm ss}(\tau_{\rm d})$ plane is presented in the inset of Fig.~\ref{fig:L_ss_vs_beta_ss}, where the time evolution of a growing pulse initially residing close to the steady-state point for $\tau_{\rm d}\=1.04\tau_{\rm min}$, follows the $\beta_{\rm ss}(L_{\rm ss})$ line arbitrarily far from the initial point. Moreover, it has been shown that the time evolution along the $\beta_{\rm ss}(L_{\rm ss})$ line is slow, in the sense that $c_s^{-1}\dot{L}(t)/\beta(t)\!\ll\!1$~\cite{pomyalov2023self}.

Our main goal is to obtain an analytic equation of motion for unsteady pulses that predicts the rate at which the pulse, represented by a point moving along a steady-state line corresponding to any pair of physical pulse observables, propagates. This equation should also demonstrate the unstable nature of steady-state pulses and provide insight into the slowness of the instability, in the sense defined above. It is this slow evolution that makes slip pulses dynamically relevant in a broad range of natural and manmade frictional systems. It is clear that the equation of motion sought for is an ordinary differential equation in terms of a single pulse observable, from which the time evolution of any other pulse observable can by obtained using the relevant steady-state relation. For example, solving for $\beta(t)$, one can readily obtain $L(t)$ using $L(t)\=L_{\rm ss}[\beta(t)]$.

An equation of motion for unsteady pulses, taking into account the properties and insights discussed above, has been recently proposed in~\cite{brantut2019stability}, which we closely follow here. We do not repeat the mathematical development, but rather focus on the physical picture and assumptions, and on the outcome. In~\cite{brantut2019stability}, a ``dislocation approximation'' of the unsteady pulse problem is adopted. Within this framework, the motion of pulses is investigated on spatial scales much larger than the instantaneous pulse size $L(t)$. On such scales, the pulse can be viewed as a dislocation whose topological charge corresponds to the instantaneous accumulated slip $b(t)$, i.e., its effective Burgers vector (hence the symbol). In more technical terms, $L(t)$ serves as an ``inner lengthscale'' such that the dislocation approximation is valid on intermediate scales $X$ that satisfy $L(t)\!\ll\!X\!\ll\!L_{\rm out}$, where $L_{\rm out}$ serves as an ``outer lengthscale''.

Invoking the leading order correction to a constantly moving dislocation field in terms of dislocation acceleration/deceleration (and hence also $\dot{b}(t)\!\ne\!0$) and the relations to steady-state pulses, the following equation of motion has been obtained~\cite{brantut2019stability}
\begin{equation}
    \frac{\Psi[b(t)]}{\beta_{\rm ss}[b(t)] c_s}\,\frac{db(t)}{dt} \simeq \frac{\tau_{\rm d}-\tau_{\rm d,ss}[b(t)]}{\mu} \ ,
    \label{eq:EOM}
\end{equation}
where $\tau_{\rm d,ss}(b)$ is the inverse of $b_{\rm ss}(\tau_{\rm d})$. The dimensionless functional $\Psi[b]$ (see Eq.~(37) in~\cite{brantut2019stability}) reads
\begin{equation}
    \Psi[b] = \frac{\left[1-\beta^2_{\rm ss}(b)\right]^{-1/4}}{2\pi}\frac{d}{db}\!\left[\frac{b}{\left[1-\beta^2_{\rm ss}(b)\right]^{1/4}}\right]\!\log\!\left[\frac{\!L_{\rm out}}{L_{\rm ss}(b)\!}\right] \ ,
    \label{eq:Psi}
\end{equation}
where $L_{\rm out}\!\gg\!L_{\rm ss}(b)$. A clear limitation here is that the outer scale $L_{\rm out}$ remains unspecified. Yet, due to the weak logarithmic dependence on $L_{\rm out}$, the equation still possess predictive powers. Specifically, for the thermal pressurization constitutive law, a satisfactory quantitative agreement with the numerically calculated exponential growth rate of perturbations near steady-state has been demonstrated using $L_{\rm out}/L_{\rm ss}(b)\!\sim\!{\cal O}(10)$.

An explicit analytic version of Eq.~\eqref{eq:EOM} has not been obtained for the thermal pressurization constitutive law and the predictions of the equation of motion far from steady-state have not been tested numerically. We aim at achieving these goals for rate-and-state friction pulses. Using the steady-state relations in Eqs.~\eqref{eq:b_ss_vs_tau_d}-\eqref{eq:L_ss_vs_beta_ss}, along with the additional relation $\beta_{\rm ss}(b_{\rm ss})\=b_{\rm ss}/\!\sqrt{\bar{\delta}^2+b_{\rm ss}^2}$ (where $\bar{\delta}\!\equiv\!2c_s G_{\rm c}/(v_0\mu)$ is a characteristic accumulated slip), we obtain~\cite{SM}
\begin{equation}
    \bar{r}_{\rm b}(b; \tau_{\rm d}) \equiv {\cal T}(b)\,\frac{\dot{b}}{b} = \frac{\displaystyle b/b_{\rm ss}(\tau_{\rm d})-1}{2 + 3\,(b/\bar{\delta})^2}\,\frac{4}{\log(\bar{\ell})} \ ,
    \label{eq:r_b_vs_b}
    \vspace{0.1cm}
\end{equation}
with ${\cal T}(b)\!\equiv\!c_s^{-1}L_{\rm ss}[b(t)]/\beta_{\rm ss}[b(t)]\=c_s^{-1}L(t)/\beta(t)$ and $\bar{\ell}\!\equiv\!L_{\rm out}/L_{\rm ss}(b)$. $\bar{r}_{\rm b}(b; \tau_{\rm d})$ is the relevant dimensionless rate of change of $b(t)$ (more generally, $\bar{r}_{\!_{\!\cal A}}(t; \tau_{\rm d})\!\equiv\!c_s^{-1}\dot{\cal A}(t)L(t)/({\cal A}(t)\beta(t))$, for any pulse observable ${\cal A}$), which quantifies the relative change of $b$ during the time ${\cal T}$ it takes the unsteady pulse to propagate its own size. Equation~\eqref{eq:r_b_vs_b} is an ordinary differential equation for $b(t)$ that is tested next.

{\em Testing the theoretical predictions}.---Eq.~\eqref{eq:r_b_vs_b} immediately predicts that the steady-state solution $b\=b_{\rm ss}(\tau_{\rm d})$ is categorically unstable for any $\tau_{\rm d}$, as observed numerically~\cite{Brener2018,brantut2019stability,pomyalov2023dynamics}. Specifically, it predicts that positive perturbations, $\delta{b}\!>\!0$, lead to growing pulses and negative ones, $\delta{b}\!<\!0$, lead to decaying pulses since all other quantities in the equation are positive. Another way of seeing it is that Eq.~\eqref{eq:EOM} predicts that if $\Psi(b)\!>\!0$ (which is the case), then the instability criterion reads $d\tau_{\rm d,ss}/db|_{b=b_{\rm ss}}\!<\!0$, as explicitly stated in~\cite{brantut2019stability}. Inverting Eq.~\eqref{eq:b_ss_vs_tau_d}, one obtains $\tau_{\rm d,ss}(b)\=\tau_{\rm res}+G_{\rm c}/b$ for the rate-and-state friction constitutive law, leading to
\begin{equation}
    d\tau_{\rm d,ss}/db |_{b=b_{\rm ss}} = -G_{\rm c}/b^{2}_{\rm ss} < 0 \ ,
    \label{eq:stability_criterion_RSF}
\end{equation}
which indeed implies a categoric instability.
\begin{figure}[htbp!]
    \centering
    \includegraphics[width=0.9\linewidth]{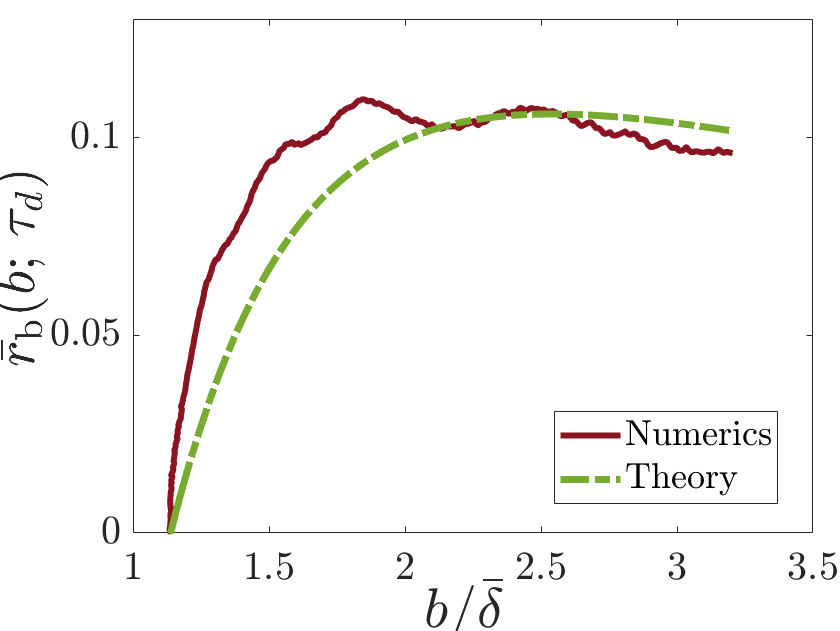}
	\caption{The theoretical prediction in Eq.~\eqref{eq:r_b_vs_b}, plotted as $\bar{r}_{\rm b}(b; \tau_{\rm d})$ vs.~$b/\bar{\delta}$ (dashed-dotted green line), compared to the numerical results of~\cite{pomyalov2023dynamics} for $\tau_{\rm d}\!=\!1.04\tau_{\rm min}$ (solid brown line). A constant $L_{\rm out}/L_{\rm ss}(b)\!=\!8.5$ is used in Eq.~\eqref{eq:r_b_vs_b} and the first data point in the theoretical prediction was shifted by $\Delta{b}$ due to the initial perturbation in the numerics (such that both vanish at the first data point).}
    \label{fig:EOM_theory_numerics}
\end{figure}

Next, we test the prediction of Eq.~\eqref{eq:r_b_vs_b} for the exponential growth/decay rate perturbed steady-state pulses. Such an analysis has been numerically performed in~\cite{pomyalov2023dynamics} for steady-state rate-and-state friction pulses with $\tau_{\rm d}\=1.03\tau_{\rm min}$ (see inset of Fig.~2A therein), finding a dimensionless growth rate of $0.13$ and a dimensionless decay rate of $0.17$. This dimensionless exponential rate can be readily obtained analytically from Eq.~\eqref{eq:r_b_vs_b} by considering $b(t\=0)\=b_{\rm ss}(\tau_{\rm d})\pm \delta{b}(t\=0)$ and finding the short-time exponential evolution of $\delta{b}(t)$~\cite{SM}. For $\tau_{\rm d}\=1.03\tau_{\rm min}$, we find a dimensionless rate of $\simeq 0.16$ once we set $\bar{\ell}\=L_{\rm out}/L_{\rm ss}(b)\=8.5$~\cite{SM}, in agreement with the numerical data of~\cite{pomyalov2023dynamics}. The selected value of $L_{\rm out}/L_{\rm ss}(b)$ is consistent with the choice in~\cite{brantut2019stability}, and is fixed hereafter.

Finally, we test the prediction of Eq.~\eqref{eq:r_b_vs_b} with no free parameters over longer timescales, when unsteady growing pulses move away from the steady-state point. We consider the growing pulse shown in the inset of Fig.~\ref{fig:L_ss_vs_beta_ss} at a fixed $\tau_{\rm d}\!=\!1.04\tau_{\rm min}$ and plot its dimensionless growth rate $\bar{r}_{\rm b}(b; \tau_{\rm d})$ in Fig.~\ref{fig:EOM_theory_numerics}. Note that $\bar{r}_{\rm b}(b; \tau_{\rm d})\!\ll\!1$ during the entire unsteady evolution, demonstrating the slowness of the instability. We then superpose the analytic prediction on the right-hand-side of Eq.~\eqref{eq:r_b_vs_b}, again using $\bar{\ell}\=L_{\rm out}/L_{\rm ss}(b)\=8.5$. The theoretical prediction is in good agreement with the numerical results --- including the initial rise, the relatively flat maximum and the beginning of the rate decline with increasing $b$ --- providing significant support to Eq.~\eqref{eq:r_b_vs_b}.

Equation~\eqref{eq:r_b_vs_b} also reveals the origin of the slowness of evolution of unsteady slip pulses, i.e., $\bar{r}_{\rm b}(b; \tau_{\rm d})\!\ll\!1$. First, the effective spatial scales separation $\bar{\ell}\=L_{\rm out}/L_{\rm ss}(b)\!\gg\!1$ contributes to it, albeit logarithmically. On top of $1/log(\bar{\ell})$, which is not much smaller than unity for $\bar{\ell}\=8.5$, there exist no other small parameters. Second, Eq.~\eqref{eq:r_b_vs_b} shows that the different powers of $b$ in the numerator and denominator control the smallness of $\bar{r}_{\rm b}(b; \tau_{\rm d})$; specifically, the numerator depends linearly on the normalized distance from the steady-state point, while the denominator grows quadratically with $b/\bar{\delta}$, which can become quite significantly larger than unity for growing pulses, see Fig.~\ref{fig:b_ss_vs_tau_d} and recall that growing pulses correspond to a decreasing $\tau_{\rm d,ss}$. In fact, Eq.~\eqref{eq:r_b_vs_b} predicts that $\bar{r}_{\rm b}(b; \tau_{\rm d})$ will eventually decrease with increasing $b$, as indeed observed in Fig.~\ref{fig:EOM_theory_numerics}.

{\em Brief summary}.---In this work, we presented and tested an analytic equation of motion for unsteady slip pulses propagating under a generic rate-and-state friction constitutive law. This progress is based on deep insights recently gained into the properties and structure of such an equation~\cite{Brener2018,brantut2019stability,pomyalov2023self,pomyalov2023dynamics}, and proceeded in two steps. First, the family of steady-state rate-and-state friction pulse solutions, numerically obtained in~\cite{pomyalov2023self}, has been effectively mapped onto the analytic ideal pulse solution~\cite{freund1979mechanics}. Second, the slip pulse equation of motion developed in~\cite{brantut2019stability} based on the ``dislocation approximation'' has been used to obtain an analytic slip pulse equation of motion, presented in Eq.~\eqref{eq:r_b_vs_b}. It is an ordinary differential equation, i.e., an equation for a single degree of freedom --- a pulse observable ---, from which the dynamics of other pulse observables can be obtained through steady-state relations, as discussed above.  

Equation~\eqref{eq:r_b_vs_b} correctly predicts the categoric instability of steady-state pulses, and quantitatively agrees with recently obtained numerical solutions for growing pulses~\cite{pomyalov2023dynamics}, both close to steady-state and far away from it. It also provides insight into the origin of the slow evolution of the instability, which explains the dynamic relevance of pulses in a broad range of natural and manmade frictional systems. That is, growing pulses are ``sustained'', in the sense that they propagate many times their characteristic size without appreciably changing their properties~\cite{brantut2019stability,pomyalov2023dynamics}. Our findings for rate-and-state frictional pulses, together with those of~\cite{brantut2019stability} for a very different interfacial constitutive law --- that of thermal pressurization of fault pore fluids --- significantly support and further substantiate the pulse equation of motion developed in~\cite{brantut2019stability}. As such, they constitute basic progress in understanding a dominant mode of rupture in frictional systems.

{\em Acknowledgements}.---We thank Anna Pomyalov for her help with the numerical data and Efim Brener for useful discussions. We acknowledge support by the Israel Science Foundation (ISF grant no.~1085/20), the Minerva Foundation (with funding from the Federal German Ministry for Education and Research), the Ben May Center for Chemical Theory and Computation, and the Harold Perlman Family. 

\clearpage

\onecolumngrid
\begin{center}
	\textbf{\large Supplemental Materials for:\\ ``An equation of motion for unsteady frictional slip pulses''}
\end{center}

\setcounter{equation}{0}
\setcounter{figure}{0}
\setcounter{section}{0}
\setcounter{table}{0}
\setcounter{page}{1}
\makeatletter
\renewcommand{\theequation}{S\arabic{equation}}
\renewcommand{\thefigure}{S\arabic{figure}}
\renewcommand{\thesubsection}{S-\arabic{subsection}}
\renewcommand{\thesection}{S-\arabic{section}}
\renewcommand{\thetable}{S-\arabic{table}}
\renewcommand*{\thepage}{S\arabic{page}}

\twocolumngrid

The goal of this brief Supplementary Materials file is to provide additional technical details about the derivation of the results appearing in the manuscript and to state the values of the parameters used in the comparison to available numerical data.

\section{E\lowercase{ffective mapping to ideal steady-state pulses}}
\label{sec:ideal_pulse}

In the manuscript, we discuss in quite detail the mapping of the family of steady-state rate-and-state friction pulse solutions, recently obtained numerically in~\cite{pomyalov2023self}, to the ideal steady-state pulse solutions of~\cite{freund1979mechanics}. Here, we provide a few additional technical details in this context.

The ideal pulse solution takes the form~\cite{freund1979mechanics}
\begin{equation}
v(\xi)= v_0 \sqrt{\frac{\tfrac{1}{2}L+\xi}{\tfrac{1}{2}L-\xi}} \ ,
\label{eqS:v_Freund}
\end{equation}
for $|\xi|\!\le\!L/2$, propagating from left to right at a velocity $c_{\rm p}\=\beta c_s$. $v_0$ can be related to the total accumulated slip $b$, given as
\begin{equation}
b=(\beta c_s)^{-1}\!\!\int_{-\tfrac{1}{2}L}^{\tfrac{1}{2}L} v(\xi)d\xi \ ,
\label{eqS:b_def}
\end{equation}
which implies
\begin{equation}
v_0 =\frac{2\beta c_s b}{\pi L} \ .
\label{eqS:v_0_b_relation}
\end{equation}
The 3 unknown --- $L$, $\beta$ and $b$ --- satisfy two relations~\cite{freund1979mechanics}
\begin{eqnarray}
    \label{eqS:Freund1}
    \pi L(\tau_{\rm d}-\tau_{\rm res})&=&\mu\,b \sqrt{1-\beta^2} \ ,\\
    \label{eqS:Freund2}
    (\tau_{\rm d}-\tau_{\rm res})\,b&=& G_{\rm c} \ .
\end{eqnarray}
Here, $\mu$ is the shear modulus, $G_{\rm c}$ is the effective fracture energy at the leading edge and $\tau_{\rm res}$ is the residual stress, all are known/given (the shear wave-speed $c_s$ is also given). The driving stress $\tau_{\rm d}$ is externally controlled. As explained in the manuscript, the relations above do {\bf not} constitute a complete solution since $v_0$ is also unknown.

As is also explained in the manuscript, the family of steady-state rate-and-state friction pulse solutions of~\cite{pomyalov2023self} can be effectively mapped onto this ideal pulse solution and at the same time provide the extra missing relation that makes the solution complete. Specifically, the rate-and-state friction constitutive relation employed in~\cite{Brener2018,Barras2019,Barras2020,pomyalov2023self,pomyalov2023dynamics,pomyalov2024} features a characteristic slip velocity $v_{\rm min}$, which sets the scale for $v_0$, and a characteristic frictional strength/stess $\tau_{\rm min}$, which sets the scale for the residual stress $\tau_{\rm res}$. The two are related through the minimum of the N-shaped steady-state friction curve $\tau_{\rm ss}(v)$, i.e., $\tau_{\rm ss}(v_{\rm min})\=\tau_{\rm min}$, extensively discussed in the literature~\cite{Bar-Sinai2014}. Once $v_0\!\simeq\!v_{\rm min}$ is known, Eq.~\eqref{eqS:v_0_b_relation} becomes an extra relation, and one can solve for the 3 unknown ($b$, $L$ and $\beta$) as
\begin{eqnarray}
    \label{eqS:b_ss_vs_tau_d}
    b_{\rm ss}(\tau_{\rm d})\!&=&\!\frac{G_{\rm c}}{\tau_{\rm d}\!-\!\tau_{\rm res}} \,\Longleftrightarrow\, \tau_{\rm d,ss}(b_{\rm ss})=\tau_{\rm res}+\frac{G_{\rm c}}{b_{\rm ss}}\ ,\\
    \label{eqS:L_ss_vs_beta_ss}
    L_{\rm ss}(\beta_{\rm ss})\!&=&\!\frac{\bar{\delta}^2}{\pi} \frac{\mu}{G_{\rm c}}\frac{\beta^2_{\rm ss}}{\sqrt{1\!-\!\beta^2_{\rm ss}}} \ ,\\
    \label{eqS:beta_ss_vs_b_ss}
    \beta_{\rm ss}(b_{\rm ss})\!&=&\!\frac{b_{\rm ss}}{\sqrt{\bar{\delta}^2\!+\!b_{\rm ss}^2}} \,\Longleftrightarrow\, b_{\rm ss}(\beta_{\rm ss})\!=\!\frac{\bar{\delta}\,\beta_{\rm ss}}{\sqrt{1\!-\!\beta^2_{\rm ss}}} \ .
\end{eqnarray}
with
\begin{equation}
    \bar{\delta} \equiv \frac{2c_s G_{\rm c}}{v_0 \mu} \ ,
    \label{eqS:bar_delta}
\end{equation}
being a characteristic accumulated slip. Equation~\eqref{eqS:b_ss_vs_tau_d}, which is identical to Eq.~(1) in the manuscript, has been compared in Fig.~2 in the manuscript to the numerical pulse solutions of~\cite{pomyalov2023self}, allowing to determine $G_{\rm c}$ and $\tau_{\rm res}$ (see values in the manuscript and in Sect.~\ref{sec:parameters}). Equation~\eqref{eqS:L_ss_vs_beta_ss}, which is identical to Eq.~(2) in the manuscript once Eq.~\eqref{eqS:bar_delta} is used, has been compared in Fig.~3 in the manuscript to the numerical pulse solutions of~\cite{pomyalov2023self}, allowing to determine $v_0$ (see value in the manuscript and in Sect.~\ref{sec:parameters}).

\section{A\lowercase{n analytic equation of motion for unsteady rate-and-state slip pulses}}
\label{sec:EOM}

An analytic equation of motion for unsteady rate-and-state slip pulses appears in Eq.~(5) in the manuscript, derived using other equations presented therein. Here, all steps in the derivation are explicitly presented. Equation~(3) in the manuscript --- the equation of motion developed in~\cite{brantut2019stability} --- takes the form
\begin{equation}
    \frac{\Psi[b(t)]}{\beta_{\rm ss}[b(t)] c_s}\,\frac{db(t)}{dt} \simeq \frac{\tau_{\rm d}-\tau_{\rm d,ss}[b(t)]}{\mu} \ ,
    \label{eqS:EOM}
\end{equation}
where the following dimensionless function
\begin{equation}
    \Psi[b] = \frac{\left[1-\beta^2_{\rm ss}(b)\right]^{-1/4}}{2\pi}\frac{d}{db}\!\left[\frac{b}{\left[1-\beta^2_{\rm ss}(b)\right]^{1/4}}\right]\!\log\!\left[\frac{\!L_{\rm out}}{L_{\rm ss}(b)\!}\right]
    \label{eqS:Psi}
\end{equation}
appears. Using Eq.~\eqref{eqS:beta_ss_vs_b_ss}, we obtain
\begin{equation}
    b\,\beta_{\rm ss}(b) \frac{d\beta_{\rm ss}(b)}{db} = \beta^2_{\rm ss}(b)\left[1-\beta^2_{\rm ss}(b) \right] \ ,
    \label{eqS:intermediate_step}
\end{equation}
which immediately implies
\begin{equation}
    \Psi(b) = \frac{2+\beta^2_{\rm ss}(b)}{4\pi\sqrt{1-\beta^2_{\rm ss}(b)}} \log\!\left[\frac{\!L_{\rm out}}{L_{\rm ss}(b)\!}\right] \ .
    \label{eqS:Psi_RSF}
\end{equation}
Equation~\eqref{eqS:EOM} can be rewritten as
\begin{equation}
   \bar{r}_{\rm b}(b; \tau_{\rm d}) \equiv \frac{L_{\rm ss}(b)}{\beta_{\rm ss}(b)\,c_s}\frac{\dot{b}}{b}=\frac{L_{\rm ss}(b)\left[\tau_{\rm d}-\tau_{\rm d,ss}(b)\right]}{\mu\,b\,\Psi(b)} \ ,
    \label{eqS:r_b_vs_b}
\end{equation}
where in the manuscript we used the notation ${\cal T}(b)\=c_s^{-1}L_{\rm ss}(b)/\beta_{\rm ss}(b)$ in Eq.~(5). Let us first discuss the numerator; using Eq.~\eqref{eqS:b_ss_vs_tau_d}, we obtain
\begin{eqnarray}
    &&L_{\rm ss}(b)\left[\tau_{\rm d}-\tau_{\rm d,ss}(b)\right]=L_{\rm ss}(b)\left[\tau_{\rm d}-\tau_{\rm res}-\frac{G_{\rm c}}{b}\right] \nonumber\\
    &&=\frac{L_{\rm ss}(b)\,G_{\rm c}}{b}\left[\frac{b}{b_{\rm ss}(\tau_{\rm d})}-1\right] \ .
    \label{eqS:step1}
\end{eqnarray}
Using then Eqs.~\eqref{eqS:Freund1}-\eqref{eqS:Freund2}, we obtain
\begin{equation}
    \frac{L_{\rm ss}(b)\,G_{\rm c}}{b} = \frac{\mu\,b\sqrt{1-\beta^2}}{\pi} \ .
    \label{eqS:step2}
\end{equation}
Considering then the denominator of Eq.~\eqref{eqS:r_b_vs_b}, along with Eq.~\eqref{eqS:Psi_RSF},~\eqref{eqS:step1} and~\eqref{eqS:step2}, we end up with
\begin{eqnarray}
     \bar{r}_{\rm b}(b; \tau_{\rm d}) =\frac{L_{\rm ss}(b)}{\beta_{\rm ss}(b)\,c_s}\frac{\dot{b}}{b}
     =\left[\frac{b}{b_{\rm ss}(\tau_{\rm d})}-1\right]\frac{1-\beta^2_{\rm ss}(b)}{2+\beta^2_{\rm ss}(b)}\frac{4}{\log(\bar{\ell})} \ , \nonumber\\
    \label{eqS:step3}
\end{eqnarray}
where we used $\bar{\ell}\=L_{\rm out}/L_{\rm ss}(b)$. Then, Eq.~\eqref{eqS:step3} transforms into Eq.~(5) in the manuscript upon using Eq.~\eqref{eqS:beta_ss_vs_b_ss}.

\section{T\lowercase{he exponential growth/decay rate of perturbations close to steady-state}}
\label{eqS:exponential_rate}

In the manuscript, we test the prediction of Eq.~(5) therein for the exponential growth/decay rate of perturbations of steady-state pulses. To this aim, we write Eq.~(5) close to steady-state at a given $\tau_{\rm d}$ as
\begin{equation}
    {\cal T}(\tau_{\rm d})\,\dot{b}(t) = \frac{\displaystyle b(t)-b_{\rm ss}(\tau_{\rm d})}{2 + 3\,\left[b_{\rm ss}(\tau_{\rm d})/\bar{\delta}\right]^2}\,\frac{4}{\log(\bar{\ell})} \ .
    \label{eqS:r_b_vs_b}
\end{equation}
Considering then small perturbations around steady-state, i.e., $\delta{b}(t)\=b(t)-b_{\rm ss}(\tau_{\rm d})$, and solving for the initial exponential deviation from steady-state, we obtain the following dimensionless exponential growth/decay rate
\begin{equation}
    \label{eqS:r_b_vs_b_exp_rate}
    {\cal T}(\tau_{\rm d})\,\frac{\delta\dot{b}(t\=0)}{|\delta{b}(t\=0)|} = \frac{\displaystyle 1}{\displaystyle 2 + 3\,\left[b_{\rm ss}(\tau_{\rm d})/\bar{\delta}\right]^2}\,\frac{4}{\log(\bar{\ell})} \ ,
    \vspace{0.1cm}
\end{equation}
where growth/decay is determined by the sign of the initial perturbation, $sgn[\delta{b}(t\=0)]$. For $\tau_{\rm d}\=1.03\tau_{\rm min}$, we have $b_{\rm ss}(\tau_{\rm d})/\bar{\delta}\=1.7785$ using Eq.~(1) and Fig.~2 in the manuscript. Plugging this value into the right-hand-side of Eq.~\eqref{eqS:r_b_vs_b_exp_rate} and using $\bar{\ell}\=8.5$, we obtain a dimensionless exponential growth/decay rate of $\simeq\!0.16$. This prediction compares well with the numerical results presented in the inset of Fig.~2A of~\cite{pomyalov2023dynamics} for $\tau_{\rm d}\=1.03\tau_{\rm min}$, finding a dimensionless exponential growth rate of $0.13$ for positive perturbations and a dimensionless exponential decay rate of $0.17$ for negative perturbations.

\vspace{0.35cm}
\section{P\lowercase{arameters used in the comparison to numerical data}}
\label{sec:parameters}

In the manuscript, various parameters are extracted and used in the comparison to available numerical data. The generic rate-and-state friction constitutive law, used in~\cite{pomyalov2023self,pomyalov2023dynamics} to obtain the numerical data to which we compared our theoretical predictions, is discussed in great detail in previous publications~\cite{BarSinai2012,Bar-Sinai2014,Brener2018,Bar-Sinai2019,Barras2019,Barras2020,pomyalov2023self,pomyalov2023dynamics,pomyalov2024}. It includes the explicit expression for the frictional strength $\tau[v(x,t),\phi(x,t)]$ and the `aging law' evolution equation for $\phi(x,t)$. The sliding fixed-point of the latter equation, which allows to express $\phi$ in terms of $v$, gives rise to the N-shaped steady-state friction curve $\tau_{\rm ss}(v)$. The latter attains a local minimum at $\tau_{\rm ss}(v_{\rm min})\=\tau_{\rm min}$, where for the used constitutive parameters we have
$\tau_{\rm min}\!=\!0.3406679$ MPa (the normal stress $\sigma$ used in~\cite{pomyalov2023self,pomyalov2023dynamics} was $\sigma\=1$ MPa) and $v_{\rm min}\=0.005971$ m/s.

In Fig.~2 in the manuscript, we compared Eq.~(1) therein --- that is identical to Eq.~\eqref{eqS:b_ss_vs_tau_d} --- to the numerical solutions of~\cite{pomyalov2023self}, obtaining $\tau_{\rm res}\!=\!1.0114\tau_{\rm min}$ and $G_{\rm c}\!=\!0.51$ J/m$^2$. The latter is in good agreement with independent estimates obtained in~\cite{Barras2020,pomyalov2023self}. These values of $\tau_{\rm min}$ and $G_{\rm c}$ are used throughout the manuscript. In Fig.~3 in the manuscript, we compared Eq.~(2) therein --- that is identical to Eq.~\eqref{eqS:L_ss_vs_beta_ss} once Eq.~\eqref{eqS:bar_delta} is used, to the numerical solutions of~\cite{pomyalov2023self}, obtaining $v_0\!=\!1.15v_{\rm min}$, used throughout the manuscript (including for determining $\bar{\delta}$ in Eq.~\eqref{eqS:bar_delta}). The linear elastic bulk parameters $\mu\!=\!9$ GPa and $c_s\!=\!2739$ m/s are also used. Finally, the normalization length $L_{\rm n}\!=\!15.2$ m is used in Fig.~3 in the manuscript (see also~\cite{pomyalov2023dynamics}).


\begin{thebibliography}{44}%
\makeatletter
\providecommand \@ifxundefined [1]{%
 \@ifx{#1\undefined}
}%
\providecommand \@ifnum [1]{%
 \ifnum #1\expandafter \@firstoftwo
 \else \expandafter \@secondoftwo
 \fi
}%
\providecommand \@ifx [1]{%
 \ifx #1\expandafter \@firstoftwo
 \else \expandafter \@secondoftwo
 \fi
}%
\providecommand \natexlab [1]{#1}%
\providecommand \enquote  [1]{``#1''}%
\providecommand \bibnamefont  [1]{#1}%
\providecommand \bibfnamefont [1]{#1}%
\providecommand \citenamefont [1]{#1}%
\providecommand \href@noop [0]{\@secondoftwo}%
\providecommand \href [0]{\begingroup \@sanitize@url \@href}%
\providecommand \@href[1]{\@@startlink{#1}\@@href}%
\providecommand \@@href[1]{\endgroup#1\@@endlink}%
\providecommand \@sanitize@url [0]{\catcode `\\12\catcode `\$12\catcode
  `\&12\catcode `\#12\catcode `\^12\catcode `\_12\catcode `\%12\relax}%
\providecommand \@@startlink[1]{}%
\providecommand \@@endlink[0]{}%
\providecommand \url  [0]{\begingroup\@sanitize@url \@url }%
\providecommand \@url [1]{\endgroup\@href {#1}{\urlprefix }}%
\providecommand \urlprefix  [0]{URL }%
\providecommand \Eprint [0]{\href }%
\providecommand \doibase [0]{https://doi.org/}%
\providecommand \selectlanguage [0]{\@gobble}%
\providecommand \bibinfo  [0]{\@secondoftwo}%
\providecommand \bibfield  [0]{\@secondoftwo}%
\providecommand \translation [1]{[#1]}%
\providecommand \BibitemOpen [0]{}%
\providecommand \bibitemStop [0]{}%
\providecommand \bibitemNoStop [0]{.\EOS\space}%
\providecommand \EOS [0]{\spacefactor3000\relax}%
\providecommand \BibitemShut  [1]{\csname bibitem#1\endcsname}%
\let\auto@bib@innerbib\@empty
\bibitem [{\citenamefont {Freund}(1979)}]{freund1979mechanics}%
  \BibitemOpen
  \bibfield  {author} {\bibinfo {author} {\bibfnamefont {L.~B.}\ \bibnamefont
  {Freund}},\ }\bibfield  {title} {\bibinfo {title} {The mechanics of dynamic
  shear crack propagation},\ }\href
  {https://agupubs.onlinelibrary.wiley.com/doi/abs/10.1029/JB084iB05p02199}
  {\bibfield  {journal} {\bibinfo  {journal} {J. Geophys. Res. Solid Earth}\
  }\textbf {\bibinfo {volume} {84}},\ \bibinfo {pages} {2199} (\bibinfo {year}
  {1979})}\BibitemShut {NoStop}%
\bibitem [{\citenamefont {Heaton}(1990)}]{Heaton1990}%
  \BibitemOpen
  \bibfield  {author} {\bibinfo {author} {\bibfnamefont {T.~H.}\ \bibnamefont
  {Heaton}},\ }\bibfield  {title} {\bibinfo {title} {{Evidence for and
  implications of self-healing pulses of slip in earthquake rupture}},\ }\href
  {https://doi.org/10.1016/0031-9201(90)90002-F} {\bibfield  {journal}
  {\bibinfo  {journal} {Phys. Earth Planet. Inter.}\ }\textbf {\bibinfo
  {volume} {64}},\ \bibinfo {pages} {1} (\bibinfo {year} {1990})}\BibitemShut
  {NoStop}%
\bibitem [{\citenamefont {Perrin}\ \emph {et~al.}(1995)\citenamefont {Perrin},
  \citenamefont {Rice},\ and\ \citenamefont {Zheng}}]{Perrin1995}%
  \BibitemOpen
  \bibfield  {author} {\bibinfo {author} {\bibfnamefont {G.}~\bibnamefont
  {Perrin}}, \bibinfo {author} {\bibfnamefont {J.~R.}\ \bibnamefont {Rice}},\
  and\ \bibinfo {author} {\bibfnamefont {G.}~\bibnamefont {Zheng}},\ }\bibfield
   {title} {\bibinfo {title} {{Self-healing slip pulse on a frictional
  surface}},\ }\href {https://doi.org/10.1016/0022-5096(95)00036-I} {\bibfield
  {journal} {\bibinfo  {journal} {J. Mech. Phys. Solids}\ }\textbf {\bibinfo
  {volume} {43}},\ \bibinfo {pages} {1461} (\bibinfo {year}
  {1995})}\BibitemShut {NoStop}%
\bibitem [{\citenamefont {Beroza}\ and\ \citenamefont
  {Mikumo}(1996)}]{Beroza_Mikumo_1996}%
  \BibitemOpen
  \bibfield  {author} {\bibinfo {author} {\bibfnamefont {G.~C.}\ \bibnamefont
  {Beroza}}\ and\ \bibinfo {author} {\bibfnamefont {T.}~\bibnamefont
  {Mikumo}},\ }\bibfield  {title} {\bibinfo {title} {Short slip duration in
  dynamic rupture in the presence of heterogeneous fault properties},\ }\href
  {https://doi.org/https://doi.org/10.1029/96JB02291} {\bibfield  {journal}
  {\bibinfo  {journal} {J. Geophys. Res. Solid Earth}\ }\textbf {\bibinfo
  {volume} {101}},\ \bibinfo {pages} {22449} (\bibinfo {year}
  {1996})}\BibitemShut {NoStop}%
\bibitem [{\citenamefont {Beeler}\ and\ \citenamefont
  {Tullis}(1996)}]{Beeler1996}%
  \BibitemOpen
  \bibfield  {author} {\bibinfo {author} {\bibfnamefont {N.~M.}\ \bibnamefont
  {Beeler}}\ and\ \bibinfo {author} {\bibfnamefont {T.~E.}\ \bibnamefont
  {Tullis}},\ }\bibfield  {title} {\bibinfo {title} {{Self-healing slip pulses
  in dynamic rupture models due to velocity-dependent strength}},\ }\href
  {https://pubs.geoscienceworld.org/ssa/bssa/article/86/4/1130/120159/self-healing-slip-pulses-in-dynamic-rupture-models}
  {\bibfield  {journal} {\bibinfo  {journal} {Bull. Seismol. Soc. Am.}\
  }\textbf {\bibinfo {volume} {86}},\ \bibinfo {pages} {1130} (\bibinfo {year}
  {1996})}\BibitemShut {NoStop}%
\bibitem [{\citenamefont {Cochard}\ and\ \citenamefont
  {Madariaga}(1996)}]{Cochard1996}%
  \BibitemOpen
  \bibfield  {author} {\bibinfo {author} {\bibfnamefont {A.}~\bibnamefont
  {Cochard}}\ and\ \bibinfo {author} {\bibfnamefont {R.}~\bibnamefont
  {Madariaga}},\ }\bibfield  {title} {\bibinfo {title} {{Complexity of
  seismicity due to highly rate-dependent friction}},\ }\href
  {https://doi.org/10.1029/96JB02095} {\bibfield  {journal} {\bibinfo
  {journal} {J. Geophys. Res. Solid Earth}\ }\textbf {\bibinfo {volume}
  {101}},\ \bibinfo {pages} {25321} (\bibinfo {year} {1996})}\BibitemShut
  {NoStop}%
\bibitem [{\citenamefont {Andrews}\ and\ \citenamefont
  {Ben-Zion}(1997)}]{Andrews1997}%
  \BibitemOpen
  \bibfield  {author} {\bibinfo {author} {\bibfnamefont {D.~J.}\ \bibnamefont
  {Andrews}}\ and\ \bibinfo {author} {\bibfnamefont {Y.}~\bibnamefont
  {Ben-Zion}},\ }\bibfield  {title} {\bibinfo {title} {{Wrinkle-like slip pulse
  on a fault between different materials}},\ }\href
  {https://doi.org/10.1029/96JB02856} {\bibfield  {journal} {\bibinfo
  {journal} {J. Geophys. Res. Solid Earth}\ }\textbf {\bibinfo {volume}
  {102}},\ \bibinfo {pages} {553} (\bibinfo {year} {1997})}\BibitemShut
  {NoStop}%
\bibitem [{\citenamefont {Zheng}\ and\ \citenamefont {Rice}(1998)}]{Zheng1998}%
  \BibitemOpen
  \bibfield  {author} {\bibinfo {author} {\bibfnamefont {G.}~\bibnamefont
  {Zheng}}\ and\ \bibinfo {author} {\bibfnamefont {J.~R.}\ \bibnamefont
  {Rice}},\ }\bibfield  {title} {\bibinfo {title} {{Conditions under which
  velocity-weakening friction allows a self-healing versus a cracklike mode of
  rupture}},\ }\href
  {https://pubs.geoscienceworld.org/ssa/bssa/article/88/6/1466/120398/conditions-under-which-velocity-weakening-friction}
  {\bibfield  {journal} {\bibinfo  {journal} {Bull. Seismol. Soc. Am.}\
  }\textbf {\bibinfo {volume} {88}},\ \bibinfo {pages} {1466} (\bibinfo {year}
  {1998})}\BibitemShut {NoStop}%
\bibitem [{\citenamefont {Nielsen}\ \emph {et~al.}(2000)\citenamefont
  {Nielsen}, \citenamefont {Carlson},\ and\ \citenamefont
  {Olsen}}]{Nielsen2000}%
  \BibitemOpen
  \bibfield  {author} {\bibinfo {author} {\bibfnamefont {S.~B.}\ \bibnamefont
  {Nielsen}}, \bibinfo {author} {\bibfnamefont {J.~M.}\ \bibnamefont
  {Carlson}},\ and\ \bibinfo {author} {\bibfnamefont {K.~B.}\ \bibnamefont
  {Olsen}},\ }\bibfield  {title} {\bibinfo {title} {{Influence of friction and
  fault geometry on earthquake rupture}},\ }\href
  {https://doi.org/10.1029/1999JB900350} {\bibfield  {journal} {\bibinfo
  {journal} {J. Geophys. Res. Solid Earth}\ }\textbf {\bibinfo {volume}
  {105}},\ \bibinfo {pages} {6069} (\bibinfo {year} {2000})}\BibitemShut
  {NoStop}%
\bibitem [{\citenamefont {Nielsen}\ and\ \citenamefont
  {Madariaga}(2003)}]{Nielsen2003}%
  \BibitemOpen
  \bibfield  {author} {\bibinfo {author} {\bibfnamefont {S.}~\bibnamefont
  {Nielsen}}\ and\ \bibinfo {author} {\bibfnamefont {R.}~\bibnamefont
  {Madariaga}},\ }\bibfield  {title} {\bibinfo {title} {{On the Self-Healing
  Fracture Mode}},\ }\href {https://doi.org/10.1785/0120020090} {\bibfield
  {journal} {\bibinfo  {journal} {Bull. Seismol. Soc. Am.}\ }\textbf {\bibinfo
  {volume} {93}},\ \bibinfo {pages} {2375} (\bibinfo {year}
  {2003})}\BibitemShut {NoStop}%
\bibitem [{\citenamefont {Brener}\ \emph {et~al.}(2005)\citenamefont {Brener},
  \citenamefont {Malinin},\ and\ \citenamefont {Marchenko}}]{Brener2005}%
  \BibitemOpen
  \bibfield  {author} {\bibinfo {author} {\bibfnamefont {E.~A.}\ \bibnamefont
  {Brener}}, \bibinfo {author} {\bibfnamefont {S.~V.}\ \bibnamefont
  {Malinin}},\ and\ \bibinfo {author} {\bibfnamefont {V.~I.}\ \bibnamefont
  {Marchenko}},\ }\bibfield  {title} {\bibinfo {title} {{Fracture and friction:
  Stick-slip motion}},\ }\href {https://doi.org/10.1140/epje/i2004-10112-3}
  {\bibfield  {journal} {\bibinfo  {journal} {Eur. Phys. J. E}\ }\textbf
  {\bibinfo {volume} {17}},\ \bibinfo {pages} {101} (\bibinfo {year}
  {2005})}\BibitemShut {NoStop}%
\bibitem [{\citenamefont {Shi}\ \emph {et~al.}(2008)\citenamefont {Shi},
  \citenamefont {Ben-Zion},\ and\ \citenamefont {Needleman}}]{Shi2008}%
  \BibitemOpen
  \bibfield  {author} {\bibinfo {author} {\bibfnamefont {Z.}~\bibnamefont
  {Shi}}, \bibinfo {author} {\bibfnamefont {Y.}~\bibnamefont {Ben-Zion}},\ and\
  \bibinfo {author} {\bibfnamefont {A.}~\bibnamefont {Needleman}},\ }\bibfield
  {title} {\bibinfo {title} {{Properties of dynamic rupture and energy
  partition in a solid with a frictional interface}},\ }\href
  {https://doi.org/10.1016/j.jmps.2007.04.006} {\bibfield  {journal} {\bibinfo
  {journal} {J. Mech. Phys. Solids}\ }\textbf {\bibinfo {volume} {56}},\
  \bibinfo {pages} {5} (\bibinfo {year} {2008})}\BibitemShut {NoStop}%
\bibitem [{\citenamefont {Rubin}\ and\ \citenamefont
  {Ampuero}(2009)}]{Rubin2009}%
  \BibitemOpen
  \bibfield  {author} {\bibinfo {author} {\bibfnamefont {A.~M.}\ \bibnamefont
  {Rubin}}\ and\ \bibinfo {author} {\bibfnamefont {J.-P.}\ \bibnamefont
  {Ampuero}},\ }\bibfield  {title} {\bibinfo {title} {{Self-similar slip pulses
  during rate-and-state earthquake nucleation}},\ }\href
  {https://doi.org/10.1029/2009JB006529} {\bibfield  {journal} {\bibinfo
  {journal} {J. Geophys. Res.}\ }\textbf {\bibinfo {volume} {114}},\ \bibinfo
  {pages} {B11305} (\bibinfo {year} {2009})}\BibitemShut {NoStop}%
\bibitem [{\citenamefont {Garagash}(2012)}]{Garagash2012}%
  \BibitemOpen
  \bibfield  {author} {\bibinfo {author} {\bibfnamefont {D.~I.}\ \bibnamefont
  {Garagash}},\ }\bibfield  {title} {\bibinfo {title} {{Seismic and aseismic
  slip pulses driven by thermal pressurization of pore fluid}},\ }\href
  {https://doi.org/10.1029/2011JB008889} {\bibfield  {journal} {\bibinfo
  {journal} {J. Geophys. Res. Solid Earth}\ }\textbf {\bibinfo {volume}
  {117}},\ \bibinfo {pages} {B04314} (\bibinfo {year} {2012})}\BibitemShut
  {NoStop}%
\bibitem [{\citenamefont {Platt}\ \emph {et~al.}(2015)\citenamefont {Platt},
  \citenamefont {Viesca},\ and\ \citenamefont {Garagash}}]{platt2015steadily}%
  \BibitemOpen
  \bibfield  {author} {\bibinfo {author} {\bibfnamefont {J.~D.}\ \bibnamefont
  {Platt}}, \bibinfo {author} {\bibfnamefont {R.~C.}\ \bibnamefont {Viesca}},\
  and\ \bibinfo {author} {\bibfnamefont {D.~I.}\ \bibnamefont {Garagash}},\
  }\bibfield  {title} {\bibinfo {title} {Steadily propagating slip pulses
  driven by thermal decomposition},\ }\href
  {https://doi.org/10.1002/2015JB012200} {\bibfield  {journal} {\bibinfo
  {journal} {Journal of Geophysical Research: Solid Earth}\ }\textbf {\bibinfo
  {volume} {120}},\ \bibinfo {pages} {6558} (\bibinfo {year}
  {2015})}\BibitemShut {NoStop}%
\bibitem [{\citenamefont {Gabriel}\ \emph {et~al.}(2012)\citenamefont
  {Gabriel}, \citenamefont {Ampuero}, \citenamefont {Dalguer},\ and\
  \citenamefont {Mai}}]{Gabriel2012}%
  \BibitemOpen
  \bibfield  {author} {\bibinfo {author} {\bibfnamefont {A.-A.}\ \bibnamefont
  {Gabriel}}, \bibinfo {author} {\bibfnamefont {J.-P.}\ \bibnamefont
  {Ampuero}}, \bibinfo {author} {\bibfnamefont {L.~A.}\ \bibnamefont
  {Dalguer}},\ and\ \bibinfo {author} {\bibfnamefont {P.~M.}\ \bibnamefont
  {Mai}},\ }\bibfield  {title} {\bibinfo {title} {{The transition of dynamic
  rupture styles in elastic media under velocity-weakening friction}},\ }\href
  {https://doi.org/10.1029/2012JB009468} {\bibfield  {journal} {\bibinfo
  {journal} {J. Geophys. Res. Solid Earth}\ }\textbf {\bibinfo {volume}
  {117}},\ \bibinfo {pages} {B09311} (\bibinfo {year} {2012})}\BibitemShut
  {NoStop}%
\bibitem [{\citenamefont {Putelat}\ \emph {et~al.}(2017)\citenamefont
  {Putelat}, \citenamefont {Dawes},\ and\ \citenamefont
  {Champneys}}]{Putelat2017}%
  \BibitemOpen
  \bibfield  {author} {\bibinfo {author} {\bibfnamefont {T.}~\bibnamefont
  {Putelat}}, \bibinfo {author} {\bibfnamefont {J.~H.}\ \bibnamefont {Dawes}},\
  and\ \bibinfo {author} {\bibfnamefont {A.~R.}\ \bibnamefont {Champneys}},\
  }\bibfield  {title} {\bibinfo {title} {{A phase-plane analysis of localized
  frictional waves}},\ }\href {https://doi.org/10.1098/rspa.2016.0606}
  {\bibfield  {journal} {\bibinfo  {journal} {Proc. R. Soc. A Math. Phys. Eng.
  Sci.}\ }\textbf {\bibinfo {volume} {473}},\ \bibinfo {pages} {20160606}
  (\bibinfo {year} {2017})}\BibitemShut {NoStop}%
\bibitem [{\citenamefont {Michel}\ \emph {et~al.}(2017)\citenamefont {Michel},
  \citenamefont {Avouac}, \citenamefont {Lapusta},\ and\ \citenamefont
  {Jiang}}]{Michel2017}%
  \BibitemOpen
  \bibfield  {author} {\bibinfo {author} {\bibfnamefont {S.}~\bibnamefont
  {Michel}}, \bibinfo {author} {\bibfnamefont {J.-P.}\ \bibnamefont {Avouac}},
  \bibinfo {author} {\bibfnamefont {N.}~\bibnamefont {Lapusta}},\ and\ \bibinfo
  {author} {\bibfnamefont {J.}~\bibnamefont {Jiang}},\ }\bibfield  {title}
  {\bibinfo {title} {{Pulse-like partial ruptures and high-frequency radiation
  at creeping-locked transition during megathrust earthquakes}},\ }\href
  {https://doi.org/10.1002/2017GL074725} {\bibfield  {journal} {\bibinfo
  {journal} {Geophys. Res. Lett.}\ }\textbf {\bibinfo {volume} {44}},\ \bibinfo
  {pages} {8345} (\bibinfo {year} {2017})}\BibitemShut {NoStop}%
\bibitem [{\citenamefont {Lambert}\ \emph {et~al.}(2021)\citenamefont
  {Lambert}, \citenamefont {Lapusta},\ and\ \citenamefont
  {Perry}}]{lambert2021propagation}%
  \BibitemOpen
  \bibfield  {author} {\bibinfo {author} {\bibfnamefont {V.}~\bibnamefont
  {Lambert}}, \bibinfo {author} {\bibfnamefont {N.}~\bibnamefont {Lapusta}},\
  and\ \bibinfo {author} {\bibfnamefont {S.}~\bibnamefont {Perry}},\ }\bibfield
   {title} {\bibinfo {title} {Propagation of large earthquakes as self-healing
  pulses or mild cracks},\ }\href
  {https://doi.org/https://doi.org/10.1038/s41586-021-03248-1} {\bibfield
  {journal} {\bibinfo  {journal} {Nature}\ }\textbf {\bibinfo {volume} {591}},\
  \bibinfo {pages} {252} (\bibinfo {year} {2021})}\BibitemShut {NoStop}%
\bibitem [{\citenamefont {Roch}\ \emph {et~al.}(2022)\citenamefont {Roch},
  \citenamefont {Brener}, \citenamefont {Molinari},\ and\ \citenamefont
  {Bouchbinder}}]{ROCH2022104607}%
  \BibitemOpen
  \bibfield  {author} {\bibinfo {author} {\bibfnamefont {T.}~\bibnamefont
  {Roch}}, \bibinfo {author} {\bibfnamefont {E.~A.}\ \bibnamefont {Brener}},
  \bibinfo {author} {\bibfnamefont {J.-F.}\ \bibnamefont {Molinari}},\ and\
  \bibinfo {author} {\bibfnamefont {E.}~\bibnamefont {Bouchbinder}},\
  }\bibfield  {title} {\bibinfo {title} {Velocity-driven frictional sliding:
  Coarsening and steady-state pulses},\ }\href
  {https://doi.org/https://doi.org/10.1016/j.jmps.2021.104607} {\bibfield
  {journal} {\bibinfo  {journal} {J. Mech. Phys. Solids}\ }\textbf {\bibinfo
  {volume} {158}},\ \bibinfo {pages} {104607} (\bibinfo {year}
  {2022})}\BibitemShut {NoStop}%
\bibitem [{\citenamefont {Galetzka}\ \emph {et~al.}(2015)\citenamefont
  {Galetzka}, \citenamefont {Melgar}, \citenamefont {Genrich}, \citenamefont
  {Geng}, \citenamefont {Owen}, \citenamefont {Lindsey}, \citenamefont {Xu},
  \citenamefont {Bock}, \citenamefont {Avouac}, \citenamefont {Adhikari} \emph
  {et~al.}}]{galetzka2015slip}%
  \BibitemOpen
  \bibfield  {author} {\bibinfo {author} {\bibfnamefont {J.}~\bibnamefont
  {Galetzka}}, \bibinfo {author} {\bibfnamefont {D.}~\bibnamefont {Melgar}},
  \bibinfo {author} {\bibfnamefont {J.~F.}\ \bibnamefont {Genrich}}, \bibinfo
  {author} {\bibfnamefont {J.}~\bibnamefont {Geng}}, \bibinfo {author}
  {\bibfnamefont {S.}~\bibnamefont {Owen}}, \bibinfo {author} {\bibfnamefont
  {E.}~\bibnamefont {Lindsey}}, \bibinfo {author} {\bibfnamefont
  {X.}~\bibnamefont {Xu}}, \bibinfo {author} {\bibfnamefont {Y.}~\bibnamefont
  {Bock}}, \bibinfo {author} {\bibfnamefont {J.-P.}\ \bibnamefont {Avouac}},
  \bibinfo {author} {\bibfnamefont {L.}~\bibnamefont {Adhikari}}, \emph
  {et~al.},\ }\bibfield  {title} {\bibinfo {title} {Slip pulse and resonance of
  the kathmandu basin during the 2015 gorkha earthquake, nepal},\ }\href
  {https://doi.org/10.1126/science.aac6383} {\bibfield  {journal} {\bibinfo
  {journal} {Science}\ }\textbf {\bibinfo {volume} {349}},\ \bibinfo {pages}
  {1091} (\bibinfo {year} {2015})}\BibitemShut {NoStop}%
\bibitem [{\citenamefont {Dunham}\ \emph {et~al.}(2011)\citenamefont {Dunham},
  \citenamefont {Belanger}, \citenamefont {Cong},\ and\ \citenamefont
  {Kozdon}}]{Dunham2011a}%
  \BibitemOpen
  \bibfield  {author} {\bibinfo {author} {\bibfnamefont {E.~M.}\ \bibnamefont
  {Dunham}}, \bibinfo {author} {\bibfnamefont {D.}~\bibnamefont {Belanger}},
  \bibinfo {author} {\bibfnamefont {L.}~\bibnamefont {Cong}},\ and\ \bibinfo
  {author} {\bibfnamefont {J.~E.}\ \bibnamefont {Kozdon}},\ }\bibfield  {title}
  {\bibinfo {title} {Earthquake ruptures with strongly rate-weakening friction
  and off-fault plasticity, part 1: Planar faults},\ }\href
  {https://doi.org/10.1785/0120100075} {\bibfield  {journal} {\bibinfo
  {journal} {Bulletin of the Seismological Society of America}\ }\textbf
  {\bibinfo {volume} {101}},\ \bibinfo {pages} {2296} (\bibinfo {year}
  {2011})}\BibitemShut {NoStop}%
\bibitem [{\citenamefont {Brener}\ \emph {et~al.}(2018)\citenamefont {Brener},
  \citenamefont {Aldam}, \citenamefont {Barras}, \citenamefont {Molinari},\
  and\ \citenamefont {Bouchbinder}}]{Brener2018}%
  \BibitemOpen
  \bibfield  {author} {\bibinfo {author} {\bibfnamefont {E.~A.}\ \bibnamefont
  {Brener}}, \bibinfo {author} {\bibfnamefont {M.}~\bibnamefont {Aldam}},
  \bibinfo {author} {\bibfnamefont {F.}~\bibnamefont {Barras}}, \bibinfo
  {author} {\bibfnamefont {J.-F.}\ \bibnamefont {Molinari}},\ and\ \bibinfo
  {author} {\bibfnamefont {E.}~\bibnamefont {Bouchbinder}},\ }\bibfield
  {title} {\bibinfo {title} {{Unstable Slip Pulses and Earthquake Nucleation as
  a Nonequilibrium First-Order Phase Transition}},\ }\href
  {https://doi.org/10.1103/PhysRevLett.121.234302} {\bibfield  {journal}
  {\bibinfo  {journal} {Phys. Rev. Lett.}\ }\textbf {\bibinfo {volume} {121}},\
  \bibinfo {pages} {234302} (\bibinfo {year} {2018})}\BibitemShut {NoStop}%
\bibitem [{\citenamefont {Pomyalov}\ \emph
  {et~al.}(2023{\natexlab{a}})\citenamefont {Pomyalov}, \citenamefont
  {Lubomirsky}, \citenamefont {Braverman}, \citenamefont {Brener},\ and\
  \citenamefont {Bouchbinder}}]{pomyalov2023self}%
  \BibitemOpen
  \bibfield  {author} {\bibinfo {author} {\bibfnamefont {A.}~\bibnamefont
  {Pomyalov}}, \bibinfo {author} {\bibfnamefont {Y.}~\bibnamefont
  {Lubomirsky}}, \bibinfo {author} {\bibfnamefont {L.}~\bibnamefont
  {Braverman}}, \bibinfo {author} {\bibfnamefont {E.~A.}\ \bibnamefont
  {Brener}},\ and\ \bibinfo {author} {\bibfnamefont {E.}~\bibnamefont
  {Bouchbinder}},\ }\bibfield  {title} {\bibinfo {title} {Self-healing
  solitonic slip pulses in frictional systems},\ }\href
  {https://doi.org/10.1103/PhysRevE.107.L013001} {\bibfield  {journal}
  {\bibinfo  {journal} {Physical Review E}\ }\textbf {\bibinfo {volume}
  {107}},\ \bibinfo {pages} {L013001} (\bibinfo {year}
  {2023}{\natexlab{a}})}\BibitemShut {NoStop}%
\bibitem [{\citenamefont {Brantut}\ \emph {et~al.}(2019)\citenamefont
  {Brantut}, \citenamefont {Garagash},\ and\ \citenamefont
  {Noda}}]{brantut2019stability}%
  \BibitemOpen
  \bibfield  {author} {\bibinfo {author} {\bibfnamefont {N.}~\bibnamefont
  {Brantut}}, \bibinfo {author} {\bibfnamefont {D.~I.}\ \bibnamefont
  {Garagash}},\ and\ \bibinfo {author} {\bibfnamefont {H.}~\bibnamefont
  {Noda}},\ }\bibfield  {title} {\bibinfo {title} {Stability of pulse-like
  earthquake ruptures},\ }\href
  {https://agupubs.onlinelibrary.wiley.com/doi/abs/10.1029/2019JB017926}
  {\bibfield  {journal} {\bibinfo  {journal} {J. Geophys. Res. Solid Earth}\
  }\textbf {\bibinfo {volume} {124}},\ \bibinfo {pages} {8998} (\bibinfo {year}
  {2019})}\BibitemShut {NoStop}%
\bibitem [{\citenamefont {Pomyalov}\ \emph
  {et~al.}(2023{\natexlab{b}})\citenamefont {Pomyalov}, \citenamefont {Barras},
  \citenamefont {Roch}, \citenamefont {Brener},\ and\ \citenamefont
  {Bouchbinder}}]{pomyalov2023dynamics}%
  \BibitemOpen
  \bibfield  {author} {\bibinfo {author} {\bibfnamefont {A.}~\bibnamefont
  {Pomyalov}}, \bibinfo {author} {\bibfnamefont {F.}~\bibnamefont {Barras}},
  \bibinfo {author} {\bibfnamefont {T.}~\bibnamefont {Roch}}, \bibinfo {author}
  {\bibfnamefont {E.~A.}\ \bibnamefont {Brener}},\ and\ \bibinfo {author}
  {\bibfnamefont {E.}~\bibnamefont {Bouchbinder}},\ }\bibfield  {title}
  {\bibinfo {title} {The dynamics of unsteady frictional slip pulses},\ }\href
  {https://doi.org/10.1073/pnas.2309374120} {\bibfield  {journal} {\bibinfo
  {journal} {Proceedings of the National Academy of Sciences}\ }\textbf
  {\bibinfo {volume} {120}},\ \bibinfo {pages} {e2309374120} (\bibinfo {year}
  {2023}{\natexlab{b}})}\BibitemShut {NoStop}%
\bibitem [{\citenamefont {Pomyalov}\ and\ \citenamefont
  {Bouchbinder}(2024)}]{pomyalov2024}%
  \BibitemOpen
  \bibfield  {author} {\bibinfo {author} {\bibfnamefont {A.}~\bibnamefont
  {Pomyalov}}\ and\ \bibinfo {author} {\bibfnamefont {E.}~\bibnamefont
  {Bouchbinder}},\ }\bibfield  {title} {\bibinfo {title} {Unsteady slip pulses
  under spatially-varying prestress},\ }\href
  {https://doi.org/https://doi.org/10.1016/j.epsl.2024.119111} {\bibfield
  {journal} {\bibinfo  {journal} {Earth and Planetary Science Letters}\
  }\textbf {\bibinfo {volume} {648}},\ \bibinfo {pages} {119111} (\bibinfo
  {year} {2024})}\BibitemShut {NoStop}%
\bibitem [{\citenamefont {Gabrieli}\ and\ \citenamefont
  {Tal}(2025)}]{gabrieli2025lab}%
  \BibitemOpen
  \bibfield  {author} {\bibinfo {author} {\bibfnamefont {T.}~\bibnamefont
  {Gabrieli}}\ and\ \bibinfo {author} {\bibfnamefont {Y.}~\bibnamefont {Tal}},\
  }\bibfield  {title} {\bibinfo {title} {Lab earthquakes reveal a wide range of
  rupture behaviors controlled by fault bends},\ }\href
  {https://doi.org/10.1073/pnas.2425471122} {\bibfield  {journal} {\bibinfo
  {journal} {Proceedings of the National Academy of Sciences}\ }\textbf
  {\bibinfo {volume} {122}},\ \bibinfo {pages} {e2425471122} (\bibinfo {year}
  {2025})}\BibitemShut {NoStop}%
\bibitem [{\citenamefont {Svetlizky}\ \emph {et~al.}(2019)\citenamefont
  {Svetlizky}, \citenamefont {Bayart},\ and\ \citenamefont
  {Fineberg}}]{Svetlizky2019}%
  \BibitemOpen
  \bibfield  {author} {\bibinfo {author} {\bibfnamefont {I.}~\bibnamefont
  {Svetlizky}}, \bibinfo {author} {\bibfnamefont {E.}~\bibnamefont {Bayart}},\
  and\ \bibinfo {author} {\bibfnamefont {J.}~\bibnamefont {Fineberg}},\
  }\bibfield  {title} {\bibinfo {title} {{Brittle Fracture Theory Describes the
  Onset of Frictional Motion}},\ }\href
  {https://doi.org/10.1146/annurev-conmatphys-031218-013327} {\bibfield
  {journal} {\bibinfo  {journal} {Annu. Rev. Condens. Matter Phys.}\ }\textbf
  {\bibinfo {volume} {10}},\ \bibinfo {pages} {031218} (\bibinfo {year}
  {2019})}\BibitemShut {NoStop}%
\bibitem [{\citenamefont {Barras}\ \emph {et~al.}(2019)\citenamefont {Barras},
  \citenamefont {Aldam}, \citenamefont {Roch}, \citenamefont {Brener},
  \citenamefont {Bouchbinder},\ and\ \citenamefont {Molinari}}]{Barras2019}%
  \BibitemOpen
  \bibfield  {author} {\bibinfo {author} {\bibfnamefont {F.}~\bibnamefont
  {Barras}}, \bibinfo {author} {\bibfnamefont {M.}~\bibnamefont {Aldam}},
  \bibinfo {author} {\bibfnamefont {T.}~\bibnamefont {Roch}}, \bibinfo {author}
  {\bibfnamefont {E.~A.}\ \bibnamefont {Brener}}, \bibinfo {author}
  {\bibfnamefont {E.}~\bibnamefont {Bouchbinder}},\ and\ \bibinfo {author}
  {\bibfnamefont {J.-F.}\ \bibnamefont {Molinari}},\ }\bibfield  {title}
  {\bibinfo {title} {Emergence of cracklike behavior of frictional rupture: The
  origin of stress drops},\ }\href {https://doi.org/10.1103/PhysRevX.9.041043}
  {\bibfield  {journal} {\bibinfo  {journal} {Phys. Rev. X}\ }\textbf {\bibinfo
  {volume} {9}},\ \bibinfo {pages} {041043} (\bibinfo {year}
  {2019})}\BibitemShut {NoStop}%
\bibitem [{\citenamefont {Barras}\ \emph {et~al.}(2020)\citenamefont {Barras},
  \citenamefont {Aldam}, \citenamefont {Roch}, \citenamefont {Brener},
  \citenamefont {Bouchbinder},\ and\ \citenamefont {Molinari}}]{Barras2020}%
  \BibitemOpen
  \bibfield  {author} {\bibinfo {author} {\bibfnamefont {F.}~\bibnamefont
  {Barras}}, \bibinfo {author} {\bibfnamefont {M.}~\bibnamefont {Aldam}},
  \bibinfo {author} {\bibfnamefont {T.}~\bibnamefont {Roch}}, \bibinfo {author}
  {\bibfnamefont {E.~A.}\ \bibnamefont {Brener}}, \bibinfo {author}
  {\bibfnamefont {E.}~\bibnamefont {Bouchbinder}},\ and\ \bibinfo {author}
  {\bibfnamefont {J.-F.}\ \bibnamefont {Molinari}},\ }\bibfield  {title}
  {\bibinfo {title} {The emergence of crack-like behavior of frictional
  rupture: Edge singularity and energy balance},\ }\href
  {https://doi.org/https://doi.org/10.1016/j.epsl.2019.115978} {\bibfield
  {journal} {\bibinfo  {journal} {Earth Planet. Sci. Lett.}\ }\textbf {\bibinfo
  {volume} {531}},\ \bibinfo {pages} {115978} (\bibinfo {year}
  {2020})}\BibitemShut {NoStop}%
\bibitem [{\citenamefont {Dieterich}(1978)}]{Dieterich1978}%
  \BibitemOpen
  \bibfield  {author} {\bibinfo {author} {\bibfnamefont {J.~H.}\ \bibnamefont
  {Dieterich}},\ }\bibfield  {title} {\bibinfo {title} {{Time-dependent
  friction and the mechanics of stick-slip}},\ }\href
  {https://doi.org/10.1007/BF00876539} {\bibfield  {journal} {\bibinfo
  {journal} {Pure Appl. Geophys.}\ }\textbf {\bibinfo {volume} {116}},\
  \bibinfo {pages} {790} (\bibinfo {year} {1978})}\BibitemShut {NoStop}%
\bibitem [{\citenamefont {Dieterich}\ and\ \citenamefont
  {Kilgore}(1994)}]{Dieterich1994a}%
  \BibitemOpen
  \bibfield  {author} {\bibinfo {author} {\bibfnamefont {J.~H.}\ \bibnamefont
  {Dieterich}}\ and\ \bibinfo {author} {\bibfnamefont {B.~D.}\ \bibnamefont
  {Kilgore}},\ }\bibfield  {title} {\bibinfo {title} {{Direct observation of
  frictional contacts: New insights for state-dependent properties}},\ }\href
  {https://doi.org/10.1007/BF00874332} {\bibfield  {journal} {\bibinfo
  {journal} {Pure Appl. Geophys.}\ }\textbf {\bibinfo {volume} {143}},\
  \bibinfo {pages} {283} (\bibinfo {year} {1994})}\BibitemShut {NoStop}%
\bibitem [{\citenamefont {Beeler}\ \emph {et~al.}(1994)\citenamefont {Beeler},
  \citenamefont {Tullis},\ and\ \citenamefont {Weeks}}]{Beeler1994}%
  \BibitemOpen
  \bibfield  {author} {\bibinfo {author} {\bibfnamefont {N.~M.}\ \bibnamefont
  {Beeler}}, \bibinfo {author} {\bibfnamefont {T.~E.}\ \bibnamefont {Tullis}},\
  and\ \bibinfo {author} {\bibfnamefont {J.~D.}\ \bibnamefont {Weeks}},\
  }\bibfield  {title} {\bibinfo {title} {The roles of time and displacement in
  the evolution effect in rock friction},\ }\href
  {https://doi.org/https://doi.org/10.1029/94GL01599} {\bibfield  {journal}
  {\bibinfo  {journal} {Geophys. Res. Lett.}\ }\textbf {\bibinfo {volume}
  {21}},\ \bibinfo {pages} {1987} (\bibinfo {year} {1994})}\BibitemShut
  {NoStop}%
\bibitem [{\citenamefont {Marone}(1998)}]{Marone1998a}%
  \BibitemOpen
  \bibfield  {author} {\bibinfo {author} {\bibfnamefont {C.}~\bibnamefont
  {Marone}},\ }\bibfield  {title} {\bibinfo {title} {{Laboratoty-derived
  friction laws and their application to seismic faulting}},\ }\href
  {https://doi.org/10.1146/annurev.earth.26.1.643} {\bibfield  {journal}
  {\bibinfo  {journal} {Annu. Rev. Earth Planet. Sci.}\ }\textbf {\bibinfo
  {volume} {26}},\ \bibinfo {pages} {643} (\bibinfo {year} {1998})}\BibitemShut
  {NoStop}%
\bibitem [{\citenamefont {Berthoud}\ \emph {et~al.}(1999)\citenamefont
  {Berthoud}, \citenamefont {Baumberger}, \citenamefont {G'Sell},\ and\
  \citenamefont {Hiver}}]{Berthoud1999}%
  \BibitemOpen
  \bibfield  {author} {\bibinfo {author} {\bibfnamefont {P.}~\bibnamefont
  {Berthoud}}, \bibinfo {author} {\bibfnamefont {T.}~\bibnamefont
  {Baumberger}}, \bibinfo {author} {\bibfnamefont {C.}~\bibnamefont {G'Sell}},\
  and\ \bibinfo {author} {\bibfnamefont {J.-M.}\ \bibnamefont {Hiver}},\
  }\bibfield  {title} {\bibinfo {title} {Physical analysis of the state- and
  rate-dependent friction law: Static friction},\ }\href
  {https://doi.org/10.1103/PhysRevB.59.14313} {\bibfield  {journal} {\bibinfo
  {journal} {Phys. Rev. B}\ }\textbf {\bibinfo {volume} {59}},\ \bibinfo
  {pages} {14313} (\bibinfo {year} {1999})}\BibitemShut {NoStop}%
\bibitem [{\citenamefont {Baumberger}\ and\ \citenamefont
  {Caroli}(2006)}]{Baumberger2006Solid}%
  \BibitemOpen
  \bibfield  {author} {\bibinfo {author} {\bibfnamefont {T.}~\bibnamefont
  {Baumberger}}\ and\ \bibinfo {author} {\bibfnamefont {C.}~\bibnamefont
  {Caroli}},\ }\bibfield  {title} {\bibinfo {title} {Solid friction from
  stick-slip down to pinning and aging},\ }\href
  {https://doi.org/https://doi.org/10.1080/00018730600732186} {\bibfield
  {journal} {\bibinfo  {journal} {Advances in Physics}\ }\textbf {\bibinfo
  {volume} {55}},\ \bibinfo {pages} {279} (\bibinfo {year} {2006})}\BibitemShut
  {NoStop}%
\bibitem [{\citenamefont {Ben-David}\ \emph {et~al.}(2010)\citenamefont
  {Ben-David}, \citenamefont {Rubinstein},\ and\ \citenamefont
  {Fineberg}}]{Ben-David2010}%
  \BibitemOpen
  \bibfield  {author} {\bibinfo {author} {\bibfnamefont {O.}~\bibnamefont
  {Ben-David}}, \bibinfo {author} {\bibfnamefont {S.~M.}\ \bibnamefont
  {Rubinstein}},\ and\ \bibinfo {author} {\bibfnamefont {J.}~\bibnamefont
  {Fineberg}},\ }\bibfield  {title} {\bibinfo {title} {{Slip-stick and the
  evolution of frictional strength}},\ }\href
  {https://doi.org/10.1038/nature08676} {\bibfield  {journal} {\bibinfo
  {journal} {Nature}\ }\textbf {\bibinfo {volume} {463}},\ \bibinfo {pages}
  {76} (\bibinfo {year} {2010})}\BibitemShut {NoStop}%
\bibitem [{\citenamefont {Rice}(2006)}]{Rice2006}%
  \BibitemOpen
  \bibfield  {author} {\bibinfo {author} {\bibfnamefont {J.~R.}\ \bibnamefont
  {Rice}},\ }\bibfield  {title} {\bibinfo {title} {{Heating and weakening of
  faults during earthquake slip}},\ }\href
  {https://doi.org/10.1029/2005JB004006} {\bibfield  {journal} {\bibinfo
  {journal} {J. Geophys. Res. Solid Earth}\ }\textbf {\bibinfo {volume}
  {111}},\ \bibinfo {pages} {B05311} (\bibinfo {year} {2006})}\BibitemShut
  {NoStop}%
\bibitem [{\citenamefont {Viesca}\ and\ \citenamefont
  {Garagash}(2015)}]{Viesca2015}%
  \BibitemOpen
  \bibfield  {author} {\bibinfo {author} {\bibfnamefont {R.~C.}\ \bibnamefont
  {Viesca}}\ and\ \bibinfo {author} {\bibfnamefont {D.~I.}\ \bibnamefont
  {Garagash}},\ }\bibfield  {title} {\bibinfo {title} {{Ubiquitous weakening of
  faults due to thermal pressurization}},\ }\href
  {https://doi.org/10.1038/ngeo2554} {\bibfield  {journal} {\bibinfo  {journal}
  {Nat. Geosci.}\ }\textbf {\bibinfo {volume} {8}},\ \bibinfo {pages} {875}
  (\bibinfo {year} {2015})}\BibitemShut {NoStop}%
\bibitem [{\citenamefont {Bar~Sinai}\ \emph {et~al.}(2012)\citenamefont
  {Bar~Sinai}, \citenamefont {Brener},\ and\ \citenamefont
  {Bouchbinder}}]{BarSinai2012}%
  \BibitemOpen
  \bibfield  {author} {\bibinfo {author} {\bibfnamefont {Y.}~\bibnamefont
  {Bar~Sinai}}, \bibinfo {author} {\bibfnamefont {E.~A.}\ \bibnamefont
  {Brener}},\ and\ \bibinfo {author} {\bibfnamefont {E.}~\bibnamefont
  {Bouchbinder}},\ }\bibfield  {title} {\bibinfo {title} {{Slow rupture of
  frictional interfaces}},\ }\href {https://doi.org/10.1029/2011GL050554}
  {\bibfield  {journal} {\bibinfo  {journal} {Geophys. Res. Lett.}\ }\textbf
  {\bibinfo {volume} {39}},\ \bibinfo {pages} {L03308} (\bibinfo {year}
  {2012})}\BibitemShut {NoStop}%
\bibitem [{\citenamefont {Bar-Sinai}\ \emph {et~al.}(2014)\citenamefont
  {Bar-Sinai}, \citenamefont {Spatschek}, \citenamefont {Brener},\ and\
  \citenamefont {Bouchbinder}}]{Bar-Sinai2014}%
  \BibitemOpen
  \bibfield  {author} {\bibinfo {author} {\bibfnamefont {Y.}~\bibnamefont
  {Bar-Sinai}}, \bibinfo {author} {\bibfnamefont {R.}~\bibnamefont
  {Spatschek}}, \bibinfo {author} {\bibfnamefont {E.~A.}\ \bibnamefont
  {Brener}},\ and\ \bibinfo {author} {\bibfnamefont {E.}~\bibnamefont
  {Bouchbinder}},\ }\bibfield  {title} {\bibinfo {title} {{On the
  velocity-strengthening behavior of dry friction}},\ }\href
  {https://doi.org/10.1002/2013JB010586} {\bibfield  {journal} {\bibinfo
  {journal} {J. Geophys. Res. Solid Earth}\ }\textbf {\bibinfo {volume}
  {119}},\ \bibinfo {pages} {1738} (\bibinfo {year} {2014})}\BibitemShut
  {NoStop}%
\bibitem [{\citenamefont {Bar-Sinai}\ \emph {et~al.}(2019)\citenamefont
  {Bar-Sinai}, \citenamefont {Aldam}, \citenamefont {Spatschek}, \citenamefont
  {Brener},\ and\ \citenamefont {Bouchbinder}}]{Bar-Sinai2019}%
  \BibitemOpen
  \bibfield  {author} {\bibinfo {author} {\bibfnamefont {Y.}~\bibnamefont
  {Bar-Sinai}}, \bibinfo {author} {\bibfnamefont {M.}~\bibnamefont {Aldam}},
  \bibinfo {author} {\bibfnamefont {R.}~\bibnamefont {Spatschek}}, \bibinfo
  {author} {\bibfnamefont {E.~A.}\ \bibnamefont {Brener}},\ and\ \bibinfo
  {author} {\bibfnamefont {E.}~\bibnamefont {Bouchbinder}},\ }\bibfield
  {title} {\bibinfo {title} {Spatiotemporal {Dynamics} of {Frictional}
  {Systems}: {The} {Interplay} of {Interfacial} {Friction} and {Bulk}
  {Elasticity}},\ }\href {https://doi.org/10.3390/lubricants7100091} {\bibfield
   {journal} {\bibinfo  {journal} {Lubricants}\ }\textbf {\bibinfo {volume}
  {7}},\ \bibinfo {pages} {91} (\bibinfo {year} {2019})}\BibitemShut {NoStop}%
\bibitem [{SM()}]{SM}%
  \BibitemOpen
  \href@noop {} {\emph {\bibinfo {title} {{See Supplemental Materials appended below for
  additional information}}}}\BibitemShut {NoStop}%
\end{thebibliography}
%

\end{document}